\documentclass[manuscript,screen,review=false]{acmart}
\AtBeginDocument{%
  }

\setcopyright{acmlicensed}
\copyrightyear{2026}
\acmYear{2026}
\acmDOI{XXXXXXX.XXXXXXX}

\PassOptionsToPackage{table}{xcolor}
\newcommand{\conditions}[1]{%
\begin{minipage}[t]{\linewidth}
\raggedright
#1
\end{minipage}%
}
\usepackage{array}
\usepackage{makecell}
\usepackage[table]{xcolor}
\usepackage{longtable}
\usepackage{pdflscape}
\usepackage{geometry}
\usepackage{booktabs}
\usepackage{multirow}
\usepackage{tabularx}
\usepackage{graphicx}
\usepackage{adjustbox}
\usepackage[table]{xcolor}
\usepackage{caption}
\usepackage{amsmath}
\usepackage{subcaption}
\newcommand{\cmark}{\ensuremath{\boldsymbol{\checkmark}}}

\newcolumntype{L}[1]{>{\raggedright\arraybackslash}p{#1}}
\newcolumntype{C}[1]{>{\centering\arraybackslash}p{#1}}

\begin{document}

\title{Potential and Challenges of Large Language Models for Reverse Engineering}

\author{Xinyu Hu}
\email{xinyu.hu3@mail.mcgill.ca}
\orcid{0000-0001-7463-7467}
\affiliation{%
  \institution{McGill University}
  \city{Montreal}
  \state{Quebec}
  \country{Canada}
}

\author{Zhiwei Fu}
\affiliation{%
  \institution{McGill University}
  \city{Montreal}
  \state{Quebec}
  \country{Canada}
}
\orcid{0000-0002-2763-6186}
\email{zhiwei.fu@mail.mcgill.ca}

\author{Shaocong Xie}
\affiliation{%
  \institution{McGill University}
  \city{Montreal}
  \state{Quebec}
  \country{Canada}
}
\orcid{0009-0006-5753-3897}
\email{shaocongxie0923@gmail.com}

\author{Steven H. H. Ding}
\affiliation{%
  \institution{McGill University}
  \city{Montreal}
  \state{Quebec}
  \country{Canada}
}
\email{steven.h.ding@mcgill.ca}
\orcid{0000-0003-4513-200X}

\author{Philippe Charland}
\affiliation{%
  \institution{Mission Critical Cyber Security Section, Defence R\&D Canada}
  \city{Valcartier}
  \state{Quebec}
  \country{Canada}
}
\email{philippe.charland@drdc-rddc.gc.ca}
\orcid{0000-0003-4051-9942}

\renewcommand{\shortauthors}{Hu et al.}

\begin{abstract}
Reverse engineering (RE) is central to cybersecurity, supporting tasks such as decompilation, deobfuscation, and security analysis. 
However, RE remains labor-intensive and expertise-demanding, as analysts often manually recover high-level semantics from low-level program representations. 
Recent advances in large language models (LLMs) provide a promising way to address these challenges through program artifact understanding, semantic reasoning, and tool-augmented problem solving. 
This potential has stimulated interest in LLM-assisted RE, but existing studies remain scattered across different tasks, targets, methodologies, and evaluation practices. 
Despite these advances, the literature still lacks a comprehensive survey that consolidates progress, systematizes technical choices, and clarifies open challenges and future opportunities. 
To fill this gap, we present a systematic survey of LLM-assisted RE, covering 48 peer-reviewed and published research articles identified through our search and selection protocol as of July 1, 2026. 
We develop a faceted taxonomy that organizes prior studies along six dimensions: task objective, analysis target, methodological approach, evaluation protocol, training scale, and data quality. 
We extract task formulations and experimental settings from existing studies to support comparison, reproducibility, and future research. 
From this review, we synthesize research gaps, characterize key challenges, and outline future directions toward more reliable, reproducible, and security-relevant applications of LLMs in RE. 


\end{abstract}



\received{7 July 2026}

\maketitle

\section{Introduction}
In modern cybersecurity, reverse engineering (RE) is indispensable for exposing hidden behaviors, implementation logic, and design rationale in programs whose source code is not accessible.
RE operates across multiple layers of abstraction: analyzing binary- and instruction-level sets, semantic reasoning about program behaviors, and also overcoming countermeasures against RE including obfuscation~\cite{liu2025llm}, packing~\cite{kim2010design}, and anti-debugging~\cite{chen2008towards}. 
Its practical scope is further broadened by the diversity of security objectives and deployment environments. 
Analysts need to examine desktop and mobile applications, firmware, IoT devices, cloud services, and proprietary platform ecosystems under different architectural, organizational, and jurisdictional constraints~\cite{ma2024one, liu2025function, rukhovich2024cad, liu2025armqwen2}. 
These capabilities enable core security tasks, including but not limited to vulnerability analysis~\cite{patir2025towards}, malware nalysis~\cite{feng2025llm,walton2024exploring}, intellectual property protection~\cite{zhang2018protecting,rouhani2019deepsigns}, and clone search~\cite{Steven_2019}. 

Despite its importance, conducting RE in practice remains highly challenging. 
Effective RE requires substantial domain-specific expertise, including knowledge of binary formats, instruction set architectures, compiler optimizations, operating-system internals, and anti-analysis mechanisms~\cite{Sami_2020,Granboulan_24}. 
These difficulties are further amplified by the unique constraints of security research, where key technical and resource components, such as datasets, research environments, toolchains, and evaluation platforms, are often difficult to construct, standardize, and share~\cite{zhou2024enhancing, yao2025hdldebugger}. 
High-quality RE data are frequently scarce, sensitive, platform-specific, or legally and ethically restricted~\cite{jiang2023nova}.
Reliable annotations often require expert manual analysis~\cite{decompiling}. 
And experimental settings may depend on fragile toolchains, protected execution environments, proprietary artifacts, or platform-dependent configurations~\cite{decompiling, YANG2026104468, xue2026record}. 
As a result, differences in data sources, analysis environments, and evaluation protocols can make results difficult to compare and reproduce across studies. 
This high expertise and resource barrier raises analysis cost, raises the process error rates, and limits scalability, thereby motivating the development of more automated, scalable, and reliable RE techniques~\cite{decompiling}.

To address these challenges, the RE community has developed rich techniques, such as interactive disassemblers, Ghidra~\cite{ghidra}, symbolic-execution platforms, S2E~\cite{chipounov2011s2e}, and representation-learning approaches, Asm2Vec~\cite{Steven_2019}. 
Although these efforts have advanced program analysis in many specific scenarios, they have not fully bridged the semantic gap between machine-level program artifacts and human-understandable program behavior. 
Therefore, RE remains heavily dependent on expert-driven interpretation and manual reasoning. 
Meanwhile, rapid advances in natural language processing (NLP) have highlighted the general-purpose representation and reasoning capabilities of neural architectures, particularly Transformer-based models~\cite{vaswani2017attention}. 
The emergence of large language model (LLM), e.g. GPT~\cite{brown2020language} and Gemini~\cite{team2023gemini}, further extends this paradigm, offering new opportunities for automating RE through contextual understanding, semantic reasoning, and tool-augmented problem solving over program artifacts~\cite{li2025sv,jelodar2026large}. 
Recent studies have shown their value in RE-related tasks such as code reconstruction, security analysis, symbol recovery, and binary understanding, which lower expertise barriers and reduce resource-intensive manual effort~\cite{hu2024degpt,siala2025towards,xie2024resym,boronat2025mdre}. 
Nevertheless, LLM-based RE remains immature, regularly suffering from hallucinations, weak reproducibility, and limited grounding in machine-level semantics, raising ongoing concerns about reliability and secure adoption~\cite{orgad2025llms, team2025gemma, shehzad2026reproducibility}. 

Early empirical studies have demonstrated the potential of LLMs for supporting RE-related security studies. 
In 2022, Pearce et al.~\cite{pearce2022pop} introduced one of the earliest empirical investigations of LLMs for RE by evaluating Codex through a ``pop quiz'' framework, showing promise for simple code-understanding tasks while, also revealing reliability limitations. 
Since then, LLMs have attracted increasing attention across cybersecurity, leading to a body of surveys and SoK studies on security applications, binary analysis, program analysis, malware analysis, and related topics~\cite{mantun2026evaluating, jelodar2026large, ghimire2025survey, ti2025towards,zhang2026survey}. 
For example, Xu et al.~\cite{xu2024large} reviewed LLM applications in security, Wang et al.~\cite{wang2025contemporary} categorized LLM-based program analysis methods, Jelodar et al.~\cite{jelodar2026large} surveyed LLM-assisted malware analysis, and Ti et al.~\cite{ti2025towards} examined RE-oriented language models. 
However, existing surveys either focus on broader cybersecurity and software analysis settings or address narrower task-specific perspectives. 
They do not provide a dedicated and systematic synthesis of LLM-based RE that jointly examines LLM methodological designs, analysis-target representations, software and hardware implementation environments, human involvement, semantic grounding, and RE-specific risks, and forward-looking research guidance. 
This gap motivates a systematic consolidation of the emerging studies on LLM-based RE.

The need for such consolidation is reinforced by the methodological fragmentation of existing LLM-based RE studies. 
Existing studies differ substantially in task formulations, representations, implementation settings, datasets, evaluation protocols, and assumptions about analyst involvement~\cite{ghimire2025survey, mantun2026evaluating, jelodar2026large}. 
Such heterogeneity makes it difficult to assess model capability independently of expert intervention, reproduce reported results, and determine whether LLMs can reliably translate low-level program artifacts into faithful, human-interpretable explanations~\cite{zhao2024explainability}. 
Therefore, a comprehensive and systematic review is needed to organize this emerging literature, clarify its methodological choices, and identify open challenges in this security-critical domain. 
This survey makes the following contributions: 

\begin{enumerate}
    \item A systematic review of LLM-based RE is provided, based on a corpus of 48 peer-reviewed publications published through July 1, 2026. 
    The scope, evolution, and maturity of this emerging research area are characterized.

    \item A six-dimensional taxonomy is proposed to organize existing studies along six axes: motivation, representation, operation, validation, magnitude, and reliability. 
    This taxonomy presents a structured framework for comparing heterogeneous LLM-based RE studies. 

    \item A fine-grained characterization of task formulations and experimental settings is conducted across the surveyed literature. 
    The analysis covers detailed analysis target, model selection, benchmark usage,  implementation environments, tool-augmentation strategies, and human involvement.

    \item Quantitative and qualitative analyses are performed to synthesize dominant research paradigms, common evaluation practices, recurring limitations, and relationships across taxonomy dimensions. 
    Key gaps are identified in semantic grounding, benchmark standardization, reproducibility, data quality, and practical reliability.
    
    \item We summarize open challenges and outline a research agenda for LLM-based RE, with attention to faithful natural language explanation, reliable evaluation, dual-use risks, and cybersecurity–NLP collaboration.
\end{enumerate}

In general, these contributions provide a structured account of LLM-based RE and clarify how existing studies can be compared, reproduced, and extended. 
They also identify methodological and practical gaps that may guide future work toward more reliable, interpretable, and deployable LLM-based RE.

\section{Preliminary}
\subsection{Background}
\textbf{RE} is the process of extracting knowledge from artifacts to deduce their design, function, and operation~\cite{chazelle1995reverse}. 
Coined in the 20th century and first applied in military and industrial contexts~\cite{baxter1997reverse,oed_reverse_engineering,eilam2005reversing}, RE became an established discipline in software engineering with the seminal definition of Chikofsky and Cross II in 1990~\cite{chikofsky1990reverse}. 
RE has played a broad role across technological eras and today covers tasks ranging from binary analysis, including disassembly and decompilation~\cite{Sami_2020,Granboulan_24}, to program understanding, such as identifying program entities and recovering control or calling graphs~\cite{Zeping_2020,tip2000scalable,rimsa2021practical}. 
At the highest level, it seeks to infer code specifications and intent, enabling vulnerability discovery~\cite{patir2025towards,ma2024one}, malware analysis~\cite{feng2025llm,walton2024exploring}, and clone search~\cite{Xu_2017,Steven_2019,Zeping_2020}. 
This progression from syntactic recovery to semantic comprehension defines RE’s central challenge. 
Modern tools combine disassemblers such as IDA Pro~\cite{eagle2011ida} and Ghidra~\cite{eagle2020ghidra} with analysis engines like Angr~\cite{shoshitaishvili2016state} and S2E~\cite{chipounov2011s2e}, automating exploit generation, vulnerability discovery, and protocol analysis~\cite{meng2024large, zhou2024enhancing}. 
Recently, machine learning, particularly LLMs, has been applied to semantic recovery, function summarization, and decompilation, signaling a shift toward data-driven RE.

\textbf{LLMs} are Transformer-based architectures trained on massive text and code corpora, reshaping NLP. 
Earlier models used statistical methods such as n-grams~\cite{guthrie2006closer}, limited by sparsity and short context. 
Word embeddings like Word2Vec~\cite{mikolov2013efficient} improved semantic representation, while RNNs~\cite{rumelhart1986learning,jordan1986attractor,werbos2002backpropagation} and LSTMs~\cite{hochreiter1997long} enabled variable-length sequence modeling, advanced by sequence-to-sequence frameworks~\cite{sutskever2014sequence}. 
The Transformer~\cite{vaswani2017attention} introduced self-attention for parallelism and long-range dependencies, overcoming RNN limits and forming the basis of modern LLMs~\cite{brown2020language,jiang2023mistral7b,dubey2024llama,guo2025deepseek}. 
Beyond architectural advances, LLM progress has been driven by training and alignment. 
Large-scale pretraining established general-purpose representations, later adapted to tasks via supervised fine-tuning~\cite{ouyang2022training} and parameter-efficient methods like LoRA~\cite{hu2022lora}. 
Alignment techniques such as RLHF~\cite{christiano2017deep} and its variants (e.g.,  PPO~\cite{schulman2017proximal}, DPO~\cite{rafailov2023direct}) further improved coherence, safety, and task alignment. 
Overall, advances in pretraining, fine-tuning, and alignment support the LLM ecosystem and enable applications in RE~\cite{hu2024degpt,tan2024llm4decompile, jiang2023nova}.


\begin{table*}[b]
\centering
\caption{Comparison with Recent Related Surveys, Reviews, and SoK Papers on LLM-Based RE}
\label{tab:survey_comparison}
\scriptsize
\setlength{\tabcolsep}{0pt}
\renewcommand{\arraystretch}{1.35}

\begin{tabular}{%
>{\centering\arraybackslash}p{0.18\textwidth}|
>{\centering\arraybackslash}p{0.1025\textwidth}|
>{\centering\arraybackslash}p{0.1100\textwidth}|
>{\centering\arraybackslash}p{0.1025\textwidth}|
>{\centering\arraybackslash}p{0.1125\textwidth}|
>{\centering\arraybackslash}p{0.1025\textwidth}|
>{\centering\arraybackslash}p{0.1025\textwidth}|
>{\centering\arraybackslash}p{0.0925\textwidth}|
>{\centering\arraybackslash}p{0.0825\textwidth}}
\toprule
\toprule
\textbf{Features $\backslash$ Research} &
\makecell[c]{\textbf{Chen et al.,} \\ \textbf{2026}\\\cite{mantun2026evaluating}} &
\makecell[c]{\textbf{Jelodar et al.,} \\ \textbf{2026}\\\cite{jelodar2026large}} &
\makecell[c]{\textbf{Zhang et al.,} \\ \textbf{2026}\\\cite{zhang2026survey}} &
\makecell[c]{\textbf{Ghimire et al.,} \\ \textbf{2025}\\\cite{ghimire2025survey}} &
\makecell[c]{\textbf{Ti et al.,} \\ \textbf{2025}\\\cite{ti2025towards}} &
\makecell[c]{\textbf{Wang et al.,} \\ \textbf{2025}\\\cite{wang2025contemporary}} &
\makecell[c]{\textbf{Xu et al.,} \\ \textbf{2025}\\\cite{xu2024large}} &
\textbf{Ours} \\
\midrule

\textbf{Recent (2024-2026)} 
& \cmark &  &  &  &  &  &  & \cmark \\

\textbf{RE-Specific} 
&  & \cmark &  & \cmark & \cmark &  &  & \cmark \\

\textbf{Taxonomy} 
& \cmark &  & \cmark & \cmark & \cmark & \cmark & \cmark & \cmark \\

\textbf{LLM Method} 
& \cmark & \cmark & \cmark &  & \cmark & \cmark & \cmark & \cmark \\

\textbf{I/O Representation} 
&  &  & \cmark &  & \cmark & \cmark &  & \cmark \\

\textbf{Evaluation \& Benchmarks} 
& \cmark & \cmark & \cmark & \cmark & \cmark & \cmark & \cmark & \cmark \\ [1.5pt]

\makecell{\textbf{Experimental Setting}\\\textbf{(Software \& Hardware)}} 
&  &  &  &  &  &  &  & \cmark \\

\textbf{Human Effort} 
& \cmark & \cmark &  & \cmark &  & \cmark & \cmark & \cmark \\

\textbf{Semantic Grounding} 
& \cmark & \cmark & \cmark & \cmark &  & \cmark & \cmark & \cmark \\ [1.5pt]

\makecell{\textbf{Risk \& Guidance}\\\textbf{(RE \& LLM)}} 
& \cmark & \cmark &  & \cmark & \cmark &  &  & \cmark \\

\bottomrule
\bottomrule
\end{tabular}
\end{table*}

\subsection{Comparison with Existing Surveys}
Table~\ref{tab:survey_comparison} positions this survey against representative recent surveys on LLMs for RE in cybersecurity. 
The comparison assesses each survey’s explicit and systematic coverage of key LLM-based RE aspects: recent literature coverage, RE-specific scope, LLM-centered methodology, RE task coverage, input or output representations, evaluation and benchmark analysis, data and experimental details, human effort, semantic grounding, and risk-oriented future guidance. 
These characteristics are selected because LLM-based RE lies at the intersection of rapidly evolving LLM techniques and a security-critical domain with heterogeneous tasks, artifacts, execution environments, and evaluation assumptions.

The comparison highlights three limitations in existing surveys. 
First, although several recent surveys were published in 2025 or 2026, they may not fully capture the latest LLM-based RE studies from 2024 to 2026, partly because peer review, publication, and indexing introduce delays in a fast-moving cross-domain research area. 
Second, most related surveys are not specifically centered on RE: some examine broader cybersecurity or program-analysis applications and only discuss RE-related topics as a subset~\cite{zhang2026survey, xu2024large, wang2025contemporary}, whereas others focus on narrower subareas, such as malware analysis, binary analysis, or RE-oriented language models~\cite{jelodar2026large, ti2025towards, ghimire2025survey}. 
Third, while many surveys discuss LLM methods and evaluation in general, they rarely provide a unified analysis of RE-specific methodological details, including analysis-target representations, benchmark construction, LLM selection, software and hardware environments, analyst involvement, and the extent to which models are grounded in machine-level semantics. 

In contrast, this survey is designed to consolidate LLM-based RE as a distinct research topic. 
Beyond summarizing existing studies, it systematically examines how LLMs are integrated into RE workflows, how program artifacts are represented as inputs and outputs, how evaluations and benchmarks are constructed, what experimental settings are reported, and how human effort contributes to reported results. 
It also analyzes semantic grounding, dual-use risks, and forward-looking guidance for future research and deployment. 
This comparison motivates the need for a dedicated and timely survey that reflects both the technical evolution of LLMs and the domain-specific requirements of RE.

\subsection{Corpus Construction and Selection Methodology}
\label{sec:corpus_selection_methodology}
This section defines the evidence base of this survey and describes the procedure used to construct the study corpus. 
We focus on peer-reviewed and formally published studies that apply LLMs to RE or to closely related security-oriented program understanding tasks that directly support RE. 
Candidate studies were collected through a combination of venue-based coverage and bibliographic search, and were then filtered through staged keyword retrieval, manual screening, and explicit inclusion and exclusion criteria. 
The resulting corpus is organized with consistent identifiers and serves as the empirical basis for the taxonomy and analyses presented in the following sections.

\textbf{Study Corpus Scope. }
We constructed the study corpus by systematically surveying publications from four flagship security conferences: ACM CCS~\cite{acmccs}, IEEE S\&P~\cite{ieeesp}, NDSS~\cite{ndss}, and USENIX Security~\cite{usenix}, and extended coverage to five leading artificial intelligence (AI) venues: AAAI~\cite{aaai}, ICLR~\cite{iclr}, ICML~\cite{icml}, IJCAI~\cite{ijcai}, and NeurIPS~\cite{neurips}, where LLM-for-RE research has recently emerged. 
To ensure broad yet controlled coverage, a staged keyword search strategy was adopted across ACM Digital Library~\cite{acmdl}, DBLP~\cite{dblp}, Scopus~\cite{scopus}, and Google Scholar~\cite{googlescholar}.

\textbf{Selection Methodology. }
We included papers published through July 1, 2026 that explicitly apply LLMs to RE. 
Candidate records were first retrieved using title-level RE-related terms, including "reverse engineer" and "RE".
The abbreviation "RE" was retained only when the record clearly referred to "reverse engineering". 
Within this candidate set, LLM-related title terms, including "LLM" and "large language model", were then used to identify studies at the intersection of LLMs and RE. 
Because relevant studies do not always mention both concepts explicitly in the title, Google Scholar was further used as a complementary search channel to capture records in which RE- and LLM-related terms appeared in the abstract, main text, metadata, or other indexed contexts. 
All candidates were manually screened, and studies were excluded if the relevant terms appeared only in references or citation snippets, or if RE was mentioned only tangentially without forming a substantive task, method, or evaluation target. 
For the final paper corpus, only peer-reviewed and formally published studies were retained, as they provide a more stable and verifiable evidence base and reduce the risk of drawing conclusions from unvetted or potentially misleading claims.

\textbf{Corpus Composition.} 
After applying the selection criteria, 48 peer-reviewed papers were retained for analysis. 
To support consistent referencing throughout the survey, each paper was assigned a unique identifier from "ID 1" to "ID 48". 
The identifiers were ordered by publication year in descending order, from the most recent studies to the earliest ones. 
Within the same publication year, papers were further ordered alphabetically by the first author's last name.

\begin{figure*}[t]
\centering
  \includegraphics[width=1\linewidth]{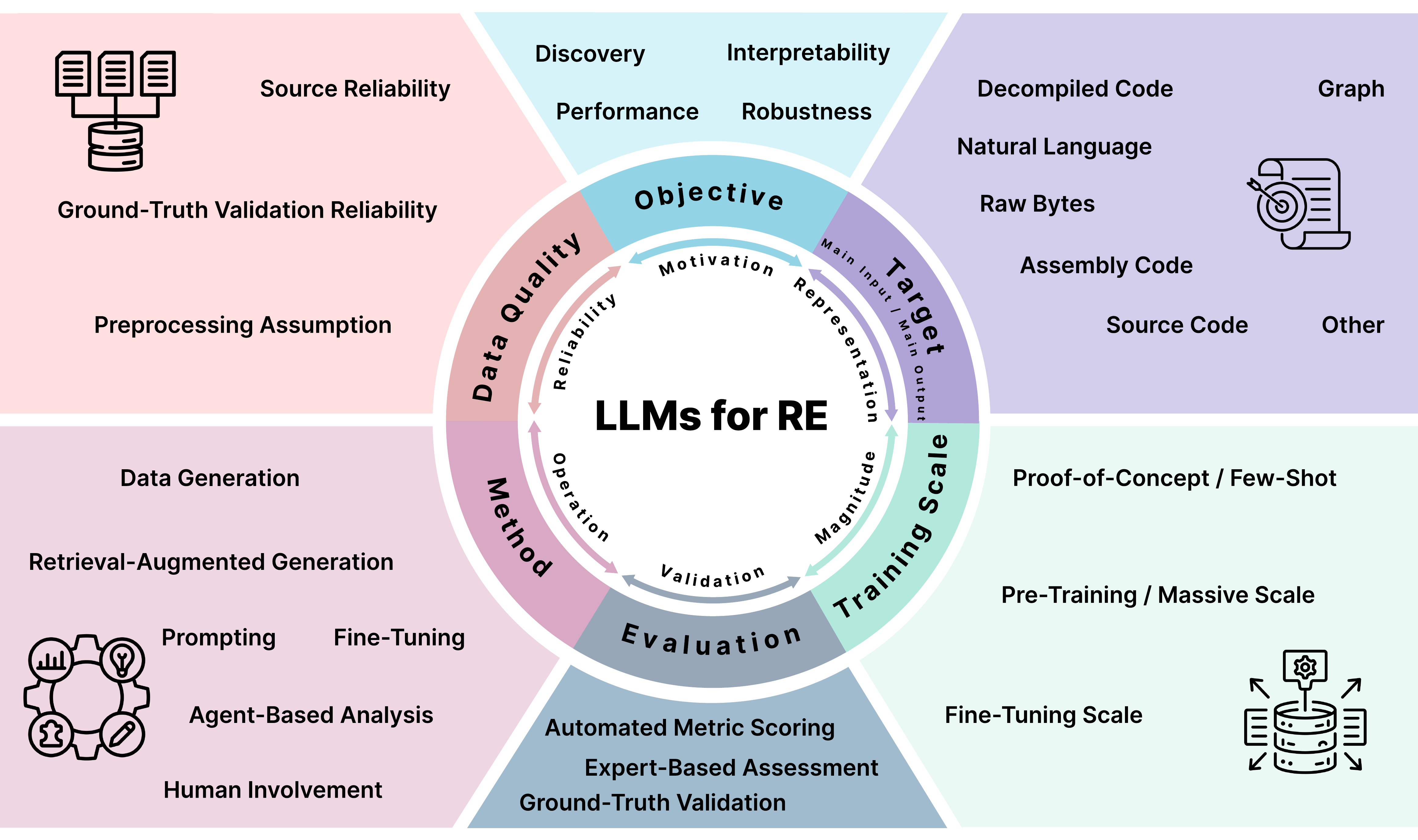}
  \caption{Overview of the proposed taxonomy, which systematizes LLM-based RE studies across six fundamental dimensions.}
  \label{fig:taxonomy}
\end{figure*}

\section{Taxonomy}
\label{sec:taxonomy}
This section introduces a faceted taxonomy for organizing the published 48 studies in the survey corpus. 
The taxonomy is organized along six complementary axes: motivation, representation, operation, validation, magnitude, and reliability. 
The motivation axis captures the primary research purpose of each study.
The representation axis describes how program artifacts are represented as model inputs and outputs. 
The operation axis characterizes how LLM-based RE methods are designed and executed. 
The validation axis summarizes how existing studies evaluate their methods. 
The magnitude axis captures the scale of empirical evidence, including dataset size, model scale, task coverage, and experimental breadth. 
The magnitude axis captures the reported or inferable scale of the LLMs used or adapted in each study. 
The reliability axis examines factors that affect trustworthiness and practical data use. 
Overall, these axes provide a structured framework for comparing heterogeneous studies, synthesizing methodological trends, and identifying research gaps in LLM-based RE. 
Figure~\ref{fig:taxonomy} illustrates the overall organization of the taxonomy.

\subsection{Dimension I: Motivation Axis - Objective}
The motivation axis is the fundamental "why" behind each study, the motivating research question or practical problem it seeks to address. 
Categorizing by objective distinguishes work that optimizes existing RE tasks from work that redefines the field’s conceptual boundaries, highlighting the overarching ambitions and priorities of the community. 
This dimension encompasses four principal attributes: performance, interpretability, discovery, and robustness. 
They form a continuum that spans from application-oriented engineering goals to knowledge-oriented exploration and systemic reliability.

\textbf{Performance} focuses on improving the efficiency or effectiveness of established well-defined RE tasks. 
Rather than introducing new capabilities, these studies seek to perform existing tasks faster, more accurately, or on a larger scale.
Performance-oriented research typically positions LLMs as automation mechanisms, emphasizing metrics such as analysis speed~\cite{chen2025suigpt}, function identification accuracy~\cite{liao2025augmenting}, and false-positive reduction~\cite{dramko2025idioms}. 
This represents the most direct use of LLMs to automate and streamline the RE pipeline.

\textbf{Interpretability} extends performance-oriented aims by focusing on making low-level, machine-oriented program representations comprehensible to human analysts.
Building upon performance-driven objectives, it addresses the challenge of bridging the semantic gap between low-level machine instructions and human-level reasoning. 
In this context, LLMs serve as semantic mediators that generate meaningful variable and function names, produce natural-language summaries, annotate program logic, and clarify complex constructs~\cite{11391932,liu2025function,petrache2025hybrid}.
Success in interpretability-oriented research is measured not by automation alone, but by the extent to which human understanding is enhanced.

\textbf{Discovery} moves beyond interpretability and automation, 
emphasizing their exploratory use to surface overlooked problems, identify underexplored methodological opportunities, and formulate hypotheses for subsequent RE investigation. 
Representative tasks include detecting unreported vulnerabilities~\cite{zhou2024enhancing}, inferring undocumented protocols~\cite{decompiling}, and uncovering latent defects~\cite{ma2024one}. 
The focus lies on generating novel insights that extend existing knowledge of software systems and reveal previously unseen behavioral or structural patterns.

\textbf{Robustness} complements prior ones by emphasizing the reliability, security, and generalizability of LLM-assisted RE methods. 
Rather than optimizing tasks, it ensures reliable operation under adversarial or uncertain conditions.
Research in this category addresses issues such as defending against prompt-injection attacks~\cite{ExposingtheGuardrails}, maintaining performance on decompiled binaries~\cite{achamyeleh2026helios, jiang2023nova}, mitigating hallucinations in safety-critical analysis~\cite{she2024wadec}, and auditing bias propagation causing security vulnerabilities.
This line of work reflects a shift toward building trust, consistency, and resilience in LLM-driven RE pipelines, while confronting whether systems remain reliable under pressure, noise, or manipulation.

\subsection{Dimension II: Representation Axis - Target}
The rpresentation axis captures each study’s “what” entities and representations analyzed in LLM-based RE. 
Target choice shapes task complexity, model capability requirements, and evaluation design, ranging from low-level binary data to high-level program representations.
Each attribute presents challenges and opportunities for RE automation. 
This dimension includes seven principal attributes: raw bytes, assembly code, decompiled code, source code, graph, natural language, and others. 
These attributes describe an abstraction spectrum from machine-level data to human-oriented representations, while also accounting for relational, explanatory, and hybrid forms of program information. 

\textbf{Raw bytes} consists of unstructured binary sequences stored on disk or in memory, representing the lowest accessible level of program data. 
Typical tasks at this target include protocol structure inference and protocol format recovery~\cite{YANG2026104468}.
For LLMs, byte-level analysis is particularly challenging, because instruction boundaries, data structures, and semantic regularities must be inferred without explicit symbolic information. 
Nevertheless, raw bytes often serve as a foundational target for subsequent disassembly, unpacking, and higher-level program reconstruction.


\textbf{Assembly code} represents the instruction-level form of machine code produced by disassemblers, where byte sequences are translated into architecture-specific mnemonics and operands. 
The challenge is to recover program semantics, control flow, data flow, and function boundaries from low-level instructions that lack source-level variables, types, and control constructs. 
LLM applications include inferring function prototypes, identifying calling conventions, detecting vulnerability patterns in instruction sequences, and summarizing the intent of assembly routines~\cite{sha2025llasm, zhou2024enhancing, petrache2025hybrid}. 
Analysis requires bridging processor-level semantics and human reasoning about program behavior. 

\textbf{Decompiled code} is source-like output reconstructed from binaries, typically as a human-readable C-like representation.
Compared with assembly, it provides explicit control structures and higher-level expressions. 
However, it remains imperfect and may contain type inconsistencies, unstructured control flow, compiler artifacts, and placeholder names such as \texttt{var1} and \texttt{var2}. 
LLMs are applied to improve interpretability, including variable and function renaming~\cite{liu2025function}, expression simplification~\cite{hu2024degpt}, logic restatement~\cite{achamyeleh2026helios}, and summary generation~\cite{kirla2025large}. 
This target provides a pragmatic middle ground between binary-level fidelity and human-oriented readability.


\textbf{Source code} denotes source-level or source-like artifacts used as high-level targets in LLM-based RE, including reference source code for evaluation and reconstructed code generated from lower-level representations~\cite{tan2024llm4decompile}. 
Unlike decompiled code, which is tied to the output quality and artifacts of a specific decompiler, source-code-level targets emphasize the recovery of human-oriented program structure, naming, and intent. 
LLMs are applied to tasks such as semantic restoration, variable renaming~\cite{xie2024resym}, control-flow reconstruction~\cite{wong2025decllm}, vulnerability detection~\cite{zhou2024enhancing}, and comment or summary generation~\cite{hatahet2025generating, app16062801}. 
This target supports higher-level program understanding, but its evaluation must account for possible gaps between source-like readability and faithful recovery of binary semantics.


\textbf{Graph} represents programs through relationships among program entities, using control-flow graphs (CFGs), call graphs, and data-flow graphs.
In these representations, nodes denote instructions, basic blocks, functions, variables, or memory objects, while edges encode control dependencies, call relations, data dependencies, or other semantic relations~\cite{xue2026record, greiner2025generalization}. 
By making relational structure explicit, graph-based representations support tasks such as binary RE visualization~\cite{brown2025large}, software architecture recovery~\cite{hatahet2025generating}, diagram recovery~\cite{greiner2025generalization, xue2026record}, and behavior modeling~\cite{ma2024one}.
For LLM-based RE, graphs provide a bridge between sequential token representations and relational program reasoning, helping models capture dependencies that are difficult to infer from raw code or binary tokens.

\textbf{Natural language} refers to textual artifacts that describe, query, explain, or summarize program behavior rather than directly encode executable logic. 
Examples include analyst questions~\cite{alazraki2026reverse}, task instructions, vulnerability descriptions~\cite{liu2025llmtaint}, malware reports~\cite{walton2024exploring}, comments~\cite{11391932}, documentation~\cite{app16062801}, function summaries~\cite{petrache2025hybrid}, behavioral rationales~\cite{walton2024exploring}, and model-generated explanations~\cite{11391932}. 
In LLM-based RE, natural language commonly appears as prompts, reasoning traces, analyst-facing outputs, or evaluation references. 
This representation is central to human-centered RE connecting low-level program artifacts with analyst-understandable descriptions. 
Its evaluation is challenging because correctness depends on faithfulness, completeness, specificity, and security relevance, not surface-level textual similarity~\cite{11391932}.

\textbf{Other} captures representations that do not fit cleanly into the preceding attributes or that combine multiple forms of program information. 
Examples may include task-specific labels~\cite{liu2025function}, debugger outputs~\cite{oliinyk2024fuzzing}, image~\cite{ExposingtheGuardrails}. 
This category is used conservatively to account for emerging or study-specific artifacts whose structure is not sufficiently standardized across the literature. 
Its presence highlights the heterogeneity of LLM-based RE inputs and the importance of explicitly reporting representation choices for comparison and reproducibility.

\subsection{Dimension III: Operation Axis - Method}


The operation axis characterizes the methodological strategies through which LLMs are applied to RE tasks. 
While the representation axis specifies what program artifacts are analyzed, this dimension describes how LLM capabilities are operationalized within RE workflows. 
This dimension consists of six attributes: prompting, fine-tuning, retrieval-augmented generation, agent-based analysis, data generation, and human involvement. 
These attributes cover different levels of model adaptation, external knowledge integration, tool use, data construction, and analyst participation. 
They reflect a methodological spectrum from lightweight use of general-purpose LLMs to specialized, workflow-aware RE systems. 
Methodological choices along this axis introduce different trade-offs among generality, task specificity, computational cost, reproducibility, interpretability, and practical performance.


\textbf{Prompting} refers to studies that apply pre-trained LLMs to RE tasks mainly through task instructions, input formatting, examples, or structured prompts, without additional model training~\cite{walton2024exploring, wong2025decllm, niang2025enhancing}. 
This category includes zero-shot, few-shot, and instruction-based prompting strategies. 
It emphasizes rapid experimentation and the reuse of general knowledge encoded in foundation models. 
Typical applications include generating labeling functions~\cite{arasteh2025trim}, summarizing binary understanding~\cite{kirla2025large}, explaining assembly instructions~\cite{petrache2025hybrid}, and translating low-level program artifacts into natural language~\cite{kirla2025large, walton2024exploring}. 
Its main advantage lies in accessibility and low computational cost, but its performance is often limited by prompt sensitivity, inconsistent outputs, and hallucination~\cite{zhuo2024prosa, anh2025survey, errica2025did}. 


\textbf{Fine-tuning} adapts LLMs to RE-specific data through additional supervised training or parameter-efficient adaptation. 
This method treats RE as a specialized domain in which generic language-model capabilities may be insufficient for accurately modeling low-level program semantics. 
Typical applications include binary-to-source decompilation~\cite{tan2024llm4decompile} and code identification~\cite{arai2025staged}, where domain alignment and prediction precision are important. 
Although fine-tuning can improve task-specific performance, it also introduces challenges related to data scarcity, annotation quality, computational cost, reproducibility, and maintenance across evolving model architectures~\cite{zheng2025learning, tan2024large, ding2023parameter, alahmari2024repeatability}.


\textbf{Retrieval-augmented generation (RAG)} enhances LLMs with external knowledge sources, such as disassembly manuals, API documentation~\cite{boronat2025mdre}, function-signature databases~\cite{chen2021sigrec}, protocol RE~\cite{YANG2026104468}, bug fixing~\cite{yao2025hdldebugger}, or domain-specific reference materials~\cite{boronat2025mdre}. 
Rather than relying solely on parametric memory, the model retrieves relevant context and incorporates it into generation or reasoning. 
This strategy can improve factual grounding and reduce hallucination, particularly in tasks that require precise knowledge of undocumented instructions, APIs, protocols, or platform-specific behaviors~\cite{YANG2026104468}. 
Its effectiveness depends on the quality, coverage, and freshness of the retrieval corpus, as well as the system's ability to retrieve relevant evidence without introducing excessive latency or irrelevant context.
    

\textbf{Agent-base analysis} conceptualizes the LLM as part of an interactive analysis workflow rather than as an isolated prediction model. 
In this setting, LLM agents may decompose RE tasks, invoke external tools, inspect intermediate outputs, revise hypotheses, and iteratively refine the analysis. 
Such workflows can integrate disassemblers, decompilers, debuggers, symbolic-execution engines, sandboxes, or other program-analysis tools, with tool outputs fed back into the LLM for further reasoning. 
This paradigm aims to approximate analyst-like investigation of complex binaries through iterative planning, tool use, and feedback~\cite{brown2025large, huang2025malpre}. 
However, it also introduces orchestration complexity, higher execution cost, weaker reproducibility, and additional security concerns, including risks associated with untrusted tool outputs and prompt-injection-style manipulation of tool-mediated contexts~\cite{brown2025large, huang2025malpre}.


\textbf{Data generation} uses LLMs to synthesize or augment data for RE tasks. 
Examples include generating synthetic code or binary-related examples~\cite{petrache2025hybrid}, producing vulnerability descriptions~\cite{oliinyk2024fuzzing}, creating task-specific labels~\cite{xie2024resym}, expanding training sets~\cite{11391932}, or generating natural-language explanations for artifacts~\cite{petrache2025hybrid}. 
This method addresses the scarcity of labeled RE data and can support downstream methods such as fine-tuning, benchmark construction, and evaluation-set expansion~\cite{jiang2023nova}. 
However, the validity of generated data remains a central concern, because incorrect, biased, or weakly grounded synthetic examples may propagate errors into downstream models and distort evaluation results~\cite{11391932}.


\textbf{Human involvement} captures methods that explicitly incorporate analyst expertise into LLM-based RE workflows. 
Such involvement may occur through expert-crafted prompts, manual annotation, feedback-based refinement, validation of model outputs, or analyst-in-the-loop reasoning~\cite{decompiling, 11391932, kirla2025large}. 
Human expertise can improve reliability, especially for tasks requiring domain knowledge, semantic judgment, or security-sensitive interpretation~\cite{chen2025suigpt, app16062801, ouf2025reverse, zhou2024enhancing}. 
At the same time, it introduces scalability constraints and complicates evaluation, because reported performance may reflect a combination of model capability and analyst expertise, rather than model behavior alone~\cite{decompiling}.

\subsection{Dimension IV: Validation Axis - Evaluation} 
The validation axis characterizes how LLM-based RE studies assess their methods’ effectiveness, reliability, and practical utility. 
This dimension consists of three attributes: expert-based assessment, automated metric scoring, and ground-truth validation. 
These attributes capture evidence used in the surveyed literature, ranging from qualitative expert judgment, to scalable metric-based evaluation and reference-based correctness checking. 
The choice of validation strategy reflects a central trade-off between scalability, semantic fidelity, reproducibility, and practical relevance. 
Automated metrics enable efficient and consistent comparison, but may fail to capture semantic correctness or analyst usability. 
Expert assessments provide richer interpretive judgment, but are costly and harder to reproduce.
Gound-truth validation offers objective reference points, but depends on the availability and quality of reliable benchmarks or labels.


\textbf{Expert-based assessment} relies on human specialists to evaluate the quality, correctness, or usefulness of LLM-generated outputs and RE-specific workflows. 
In LLM-based RE, assessment may involve judging the readability of decompiled code, the plausibility of vulnerability explanations, the accuracy of recovered semantics, or the clarity of generated summaries~\cite{hu2024degpt, yao2025hdldebugger, su2025disco, tan2024llm4decompile}. 
Its strength is that expert judgment can account for contextual factors that are difficult to capture with automatic metrics, including analyst usability, security relevance, and semantic adequacy~\cite{kirla2025large, hatahet2025generating}. 
However, this approach is resource-intensive, potentially subjective, and difficult to reproduce unless the evaluation criteria, annotator expertise, agreement procedure, and validation process are reported~\cite{hu2024degpt, alazraki2026reverse, kirla2025large}. 
Despite limitations, expert-based assessment remains important for evaluating interpretability-oriented and security-sensitive RE tasks.

\textbf{Automated metric scoring} evaluates LLM-based RE outputs using quantitative measures, including metrics adapted from NLP, software engineering, and security analysis. 
Examples include BLEU~\cite{papineni2002bleu}, ROUGE~\cite{lin2004rouge}, accuracy, precision, recall, and F1 score~\cite{lewis1995evaluating,xu2023lmpa, sha2025hyres, siala2025towards}. 
This approach enables scalable comparison across models, prompts, or system variants. 
However, automated metrics can be weak proxies for RE-specific correctness: textual similarity may not reflect functional equivalence, semantic faithfulness, vulnerability validity, or the practical usefulness of an explanation to analysts~\cite{achamyeleh2026helios}. 
Consequently, studies that rely solely on automated metrics may overestimate model effectiveness, particularly for tasks that require machine-level semantic grounding or human-interpretable reasoning.

\textbf{Ground-truth validation} assesses model outputs against authoritative references, curated labels, or reproducible benchmark outcomes. 
Examples include checking whether reconstructed code preserves original binary behavior, whether recovered symbols or function signatures match reference annotations, whether vulnerability reports match Common Vulnerabilities and Exposures entries, or whether generated patches and decompilation outputs pass correctness tests~\cite{petrache2025hybrid, liu2025llmtaint, wong2025decllm}. 
This strategy provides a rigorous basis for evaluation and comparison~\cite{achamyeleh2026helios, faruque2025gate, tan2024llm4decompile}. 
Nevertheless, its applicability is limited by the scarcity of high-quality ground truth in RE, expert annotation cost, the sensitivity of security datasets, and the difficulty of defining reference answers for open-ended tasks, such as novel vulnerability discovery or behavioral interpretation~\cite{liu2025llmtaint, huang2025malpre, YANG2026104468}. 
Thus, ground-truth validation is most effective when paired with transparent dataset construction, clearly specified evaluation protocols, and where necessary, expert review.

\subsection{Dimension V: Magnitude Axis - Training Scale}
The magnitude axis characterizes the training and adaptation scale of LLM-based RE studies. 
Whereas the operation axis describes the strategy used to apply LLMs, this dimension focuses on model training, task-specific adaptation, and computational investment required to support the reported RE capability. 
It consists of three attributes: proof-of-concept/few-shot scale, fine-tuning scale, and pretraining/massive scale. 
These attributes distinguish lightweight use of general-purpose models, domain adaptation on RE-specific data, and large-scale training or continued pretraining over code, binary, or security-related corpora. 
The magnitude axis highlights trade-offs among accessibility, specialization, computational cost, reproducibility, and generalization, while also exposing structural limitations caused by opaque training data, proprietary model development, and underreported training configurations~\cite{datascale1,datascale2}.


\textbf{Proof-of-concept/few-shot scale} refers to studies that use off-the-shelf LLMs with no additional model training or only a small number of examples provided through prompts. 
At this scale, LLMs are mainly used to demonstrate feasibility, explore task formulations, or test whether general-purpose models can perform basic RE reasoning under limited supervision~\cite{achamyeleh2026helios, hatahet2025generating, ouf2025reverse}. 
This setting enables rapid experimentation and has low computational requirements, making it useful for early-stage exploration. 
However, conclusions drawn at this scale are often limited by narrow test cases, prompt sensitivity, unstable outputs, and weak evidence of generalizability~\cite{brown2025large, liao2025augmenting}. 
Accordingly, studies in this category provide preliminary evidence of potential rather than robust validation of RE capability.


\textbf{Fine-tuning scale} refers to studies that adapt pre-trained LLMs using RE-specific training data through supervised fine-tuning, instruction tuning, or parameter-efficient adaptation. 
This scale reflects a stronger commitment to domain specialization, as model parameters or adaptation modules are updated to better capture task-specific patterns in binaries, assembly, decompiled code, vulnerability labels, or other RE artifacts. 
Typical applications include binary-to-source decompilation~\cite{tan2024llm4decompile} and bug fixing~\cite{yao2025hdldebugger}, where domain alignment and prediction precision are important. 
Fine-tuning can improve performance over direct prompting, but it also depends on the availability, quality, and representativeness of RE-specific training data~\cite{tan2024llm4decompile}. 
It may further introduce reproducibility challenges when datasets, training recipes, hyperparameters, or model checkpoints are not fully reported~\cite{rukhovich2024cad}. 
Studies at this scale therefore occupy an intermediate position between lightweight feasibility demonstrations and large-scale model development.


\textbf{Pretraining/massive scale} refers to studies or models that rely on large-scale pretraining, continued pretraining, or training from scratch over extensive corpora relevant to code, binaries, assembly, decompiled programs, vulnerability information, or security documentation~\cite{jiang2023nova}. 
At this scale, the objective is to acquire broad representations that can support transfer, zero-shot generalization, or adaptation across RE tasks and program artifacts. 
Such models may provide stronger foundations for complex RE reasoning, especially when exposed to diverse programming languages, architectures, compilation settings, or security-relevant patterns. 
However, large-scale training requires computational resources and is often associated with incomplete reporting of training data, filtering rules, compute budgets, and model-development procedures~\cite{jiang2023nova}. 
These factors limit reproducibility and make it difficult to determine how much reported RE capability comes from model scale, training data coverage, task-specific adaptation, or post-training alignment. 
This category therefore highlights both potential and transparency challenges of scaling LLMs for RE.

\subsection{Dimension VI: Reliability Axis - Data Quality}

The reliability axis evaluates the quality and trustworthiness of datasets used in LLM-based RE studies. 
Whereas the preceding dimensions characterize research objectives, targets, methods, evaluation strategies, and training scale, this dimension focuses on whether the empirical evidence supporting each study can be verified, reproduced, and compared. 
This dimension consists of three attributes: source reliability, preprocessing assumption, and ground-truth validation reliability. 
These attributes assess dataset provenance and accessibility, transparency of preprocessing procedures, and the validity of labels or ground-truth references used for training and evaluation. 
Each attribute is coded using predefined reliability levels, with the following criteria used to distinguish high- and medium-quality evidence.


\textbf{Source reliability} captures the credibility, accessibility, and traceability of datasets used in LLM-based RE studies. 
For this attribute, a dataset is rated according to the strongest applicable evidence level.
In particular, a high or medium rating is assigned when at least one criterion in the corresponding level is satisfied. 
A dataset is rated \textbf{high} if it satisfies at least one of the following conditions: 
(1) it has been cited or used in papers published at leading venues, such as ACM- or IEEE-sponsored conferences; 
(2) it is widely used and publicly hosted, for example on GitHub with more than 100 stars or on Hugging Face with more than 100 downloads; or 
(3) its source is fully described and accompanied by an access link that enables independent verification. 
A dataset is rated \textbf{medium} if it satisfies at least one of the following conditions, but does not meet any high criterion: 
(1) it has not been cited or used in papers published at leading venues; 
(2) it is described by the authors, but no access link is provided for verification; or 
(3) it is publicly hosted, but has limited community adoption, such as fewer than 100 stars on GitHub or fewer than 100 downloads on Hugging Face. 
This attribute therefore assesses whether the dataset can be independently inspected, whether its provenance is documented, and whether it has gained external validation through publication use, public availability, or community adoption.

\textbf{Preprocessing assumption} captures the transparency of transforming raw artifacts into model-ready representations. 
Such preprocessing may include disassembly, decompilation, tokenization, filtering, normalization, label extraction, feature construction, or conversion into intermediate representations~\cite{jiang2023nova, she2024wadec, xie2024resym}. 
These steps encode methodological assumptions about which program information is preserved, abstracted, or discarded before model training or evaluation~\cite{preprocessingassumptions,preprocessingassumptions2}. 
Unlike the other two reliability attributes, each level of preprocessing assumption is defined by a single criterion. 
A study is rated \textbf{high} when the authors provide both the preprocessing code and the parameter settings used in the preprocessing steps, enabling independent replication of the transformed data. 
A study is rated \textbf{medium} when the preprocessing steps are described in detail, but the corresponding code or parameter settings are not provided. 
This attribute therefore assesses whether reported results depend on transparent and reproducible preprocessing choices, or on data transformation procedures that cannot be fully reconstructed from the paper.


\textbf{Ground-truth validation reliability} captures the clarity, correctness, and verifiability of RE task labels, such as vulnerability categories, malware families, function boundaries, semantic tags, symbol names, clone labels, or correctness labels for recovered program elements~\cite{xue2026record, rukhovich2024cad, su2025disco}. 
Reliable labels are essential because ground-truth errors can propagate into training, benchmarking, and evaluation, leading to distorted claims about model performance~\cite{zhang2025rebert, tan2024llm4decompile}. 
For this attribute, label reliability is rated according to the strongest applicable evidence level; in particular, a high or medium rating is assigned when at least one criterion in the corresponding level is satisfied. 
A dataset is rated \textbf{high} if it satisfies at least one of the following conditions: 
(1) the source code is compiled into lower-level representations, such as assembly, bytecode, or binaries, for model training or evaluation, and both the original source code and the compiled artifacts are provided; 
(2) the dataset labels are clearly specified and accompanied by detailed descriptions; 
(3) the authors propose and conduct explicit ground-truth label validation for the dataset; or 
(4) the dataset labels involve expert ground-truth validation. 
A dataset is rated \textbf{medium} if it satisfies at least one of the following conditions, but does not meet any high criterion: 
(1) the original source code is not provided, but label descriptions are included; 
(2) the dataset labels are clearly specified, but not described in detail; or 
(3) the authors do not propose or conduct ground-truth label validation, but provide detailed descriptions of the labels. 
This attribute therefore assesses whether the labels used in LLM-based RE studies provide a dependable basis for training, evaluation, and cross-study comparison.

\section{Analysis of LLM-based RE Studies}
\label{sec:analysis}
Based on the taxonomy, this section analyzes the 48-study corpus from four complementary perspectives. 
First, the quantitative landscape summarizes the distribution of studies across time, representations, methodologies, strategies, scales, and reliability levels. 
Second, the intra-dimensional analysis examines how attributes within each taxonomy axis relate to one another and what trade-offs they reveal. 
Third, the cross-dimensional analysis investigates interactions among the six axes, highlighting how research objectives, analysis representations, operational methods, validation strategies, training scale, and data quality jointly shape LLM-based RE workflow. 
Finally, qualitative insights is used to derive higher-level insights about methodological maturity, reproducibility, semantic grounding, and human involvement.

\begingroup
\tiny
\setlength{\tabcolsep}{1pt}
\renewcommand{\arraystretch}{1.25}
\setlength\LTleft{0pt}
\setlength\LTright{0pt}

\begin{longtable}{|
>{\centering\arraybackslash}p{0.015\textwidth}|
>{\raggedright\arraybackslash}p{0.130\textwidth}|
>{\centering\arraybackslash}p{0.095\textwidth}|
>{\centering\arraybackslash}p{0.1425\textwidth}|
>{\centering\arraybackslash}p{0.1425\textwidth}|
>{\centering\arraybackslash}p{0.10\textwidth}|
>{\centering\arraybackslash}p{0.070\textwidth}|
>{\centering\arraybackslash}p{0.060\textwidth}|
>{\centering\arraybackslash}p{0.060\textwidth}|
>{\centering\arraybackslash}p{0.060\textwidth}|
>{\centering\arraybackslash}p{0.060\textwidth}|}

\caption{Overview of 48 published studies on LLM-based RE, organized across six dimensions: objectives (motivation), targets with main input and main output subcolumns (representation), methods (operation), evaluation (validation), training scale (magnitude), and data quality (reliability). Author-year references are shown with corresponding numbered in-text citations. Abbreviations: Human Involv. = human involvement; Agent-Based A. = agent-based analysis; RAG = retrieval-augmented generation; AMS = automated metric scoring; E-BA = expert-based assessment; GTV = ground-truth validation; PC/FS = proof-of-concept/few-shot scale; FTS = fine-tuning scale; PT/MS = pretraining/massive scale; GTV Reliab. = ground-truth validation reliability; Preproc. Assump. = preprocessing assumptions; Source Reliab. = source reliability; H = high; M = medium; “–” = not reported. 
}
\label{tab:llm-re-classification} \\

\hline
\rowcolor[HTML]{F6F8F9}
\textbf{ID} & \multicolumn{1}{c|}{\textbf{Reference}} & \textbf{Objective} &
\multicolumn{2}{c|}{\textbf{Target}} &
\textbf{Method} &
\textbf{Evaluation} &
\makecell[c]{\textbf{Training}\\ \textbf{Scale}} &
\multicolumn{3}{c|}{\textbf{Data Quality}} \\ \hline

\rowcolor[HTML]{F6F8F9}
 &  &  &
\textbf{Main Input} &
\textbf{Main Output} &
 &  &
 & \makecell[c]{\textbf{GTV}\\ \textbf{Reliab.}} &
\makecell[c]{\textbf{Preproc.}\\ \textbf{Assump.}} &
\makecell[c]{\textbf{Source}\\ \textbf{Reliab.}} \\ \hline
\endfirsthead

\hline
\rowcolor[HTML]{F6F8F9}
\textbf{ID} & \multicolumn{1}{c|}{\textbf{Reference}} & \textbf{Objective} &
\multicolumn{2}{c|}{\textbf{Target}} &
\textbf{Method} &
\textbf{Evaluation} &
\makecell[c]{\textbf{Training}\\ \textbf{Scale}} &
\multicolumn{3}{c|}{\textbf{Data Quality}} \\ \hline

\rowcolor[HTML]{F6F8F9}
 &  &  &
\textbf{Main Input} &
\textbf{Main Output} &
 &   &
 & \makecell[c]{\textbf{GTV}\\ \textbf{Reliability}} &
\makecell[c]{\textbf{Preproc.}\\ \textbf{Assump.}} &
\makecell[c]{\textbf{Source}\\ \textbf{Reliability}} \\ \hline
\endhead

\endfoot

\rowcolor[HTML]{FFFFFF}
1 & Achamyeleh et al., 2026 \cite{achamyeleh2026helios} & \makecell[c]{\textbf{Performance}\\ Robustness} & \makecell[c]{Decompiled Code \\ Graph} & Source Code & \makecell[c]{Data Generation \\ Prompting} & \makecell[c]{AMS \\ GTV} & PC/FS & H & M & H \\ \hline

\rowcolor[HTML]{F6F8F9}
2 & Basque et al., 2026 \cite{decompiling} & \makecell[c]{\textbf{Discovery} \\ Interpretability \\ Performance} & Decompiled Code & \makecell[c]{Natural Language \\ Other (recovered artifacts)} & \makecell[c]{Human Involv. \\ Prompting} & \makecell[c]{AMS \\ E-BA \\ GTV} & PC/FS & H & - & H \\ \hline

\rowcolor[HTML]{FFFFFF}
3 & Sha et al., 2026 \cite{sha2025hyres} & \makecell[c]{Interpretability \\ \textbf{Performance}} & Decompiled Code & Natural Language & Prompting & AMS & PC/FS & M & M & H \\ \hline

\rowcolor[HTML]{F6F8F9}
4 & Surk et al., 2026 \cite{app16062801} & \makecell[c]{\textbf{Discovery} \\ Interpretability \\ Performance} & Source Code & Natural Language & \makecell[c]{Human Involv. \\ Prompting} & AMS & PC/FS & M & H & M \\ \hline

\rowcolor[HTML]{FFFFFF}
5 & Xue et al., 2026 \cite{xue2026record} & \makecell[c]{Discovery \\ Interpretability \\ \textbf{Performance}} & Source Code & Graph & \makecell[c]{Agent-Based A. \\ Data Generation \\ Fine-Tuning \\ Human Involv. \\ Prompting} & \makecell[c]{AMS \\ E-BA \\ GTV} & FTS & H & M & M \\ \hline

\rowcolor[HTML]{F6F8F9}
6 & Yang et al., 2026 \cite{YANG2026104468} & \makecell[c]{\textbf{Performance} \\ Robustness} & Raw Bytes & \makecell[c]{Natural Language \\ Other (protocol structure)} & \makecell[c]{RAG \\ Prompting} & \makecell[c]{AMS \\ E-BA \\ GTV} & PC/FS & H & M & M \\ \hline

\rowcolor[HTML]{FFFFFF}
7 & Alazraki et al., 2025 \cite{alazraki2026reverse} & \makecell[c]{Discovery \\ \textbf{Robustness}} & Natural Language & Natural Language & Fine-Tuning & \makecell[c]{AMS \\ E-BA \\ GTV} & FTS & M & - & H \\ \hline

\rowcolor[HTML]{F6F8F9}
8 & Arai et al., 2025 \cite{arai2025staged} & \makecell[c]{\textbf{Performance} \\ Robustness} & Assembly Code & Assembly Code & \makecell[c]{Data Generation \\ Fine-Tuning} & \makecell[c]{AMS \\ GTV} & FTS & M & M & M \\ \hline

\rowcolor[HTML]{FFFFFF}
9 & Arasteh et al., 2025 \cite{arasteh2025trim} & \makecell[c]{\textbf{Discovery} \\ Interpretability \\ Performance} & Decompiled Code & Other (module categories) & \makecell[c]{Data Generation \\ Prompting} & \makecell[c]{AMS \\ GTV} & PC/FS & H & M & H \\ \hline

\rowcolor[HTML]{F6F8F9}
10 & Boronat et al., 2025 \cite{boronat2025mdre} & \makecell[c]{Interpretability \\ \textbf{Performance}} & Source Code & Other (domain model) & \makecell[c]{Prompting \\ RAG} & \makecell[c]{AMS \\ GTV} & PC/FS & H & M & H \\ \hline

\rowcolor[HTML]{FFFFFF}
11 & Brown et al., 2025 \cite{brown2025large} & \makecell[c]{\textbf{Discovery} \\ Interpretability} & Decompiled Code & Graph & \makecell[c]{Agent-Based A. \\ Prompting} & \makecell[c]{AMS \\ E-BA \\ GTV} & PC/FS & H & - & M \\ \hline

\rowcolor[HTML]{F6F8F9}
12 & Chen et al., 2025 \cite{chen2025suigpt} & \textbf{Performance} & Decompiled Code & Source Code & \makecell[c]{Prompting \\ Human Involv.} & \makecell[c]{AMS \\ E-BA \\ GTV} & PC/FS & H & M & M \\ \hline

\rowcolor[HTML]{FFFFFF}
13 & Choi et al., 2025 \cite{choi2024chatdeob} & \textbf{Performance} & Source Code & Source Code & \makecell[c]{Data Generation \\ Fine-Tuning \\ Prompting} & \makecell[c]{AMS \\ GTV} & FTS & H & M & M \\ \hline

\rowcolor[HTML]{F6F8F9}
14 & Dramko et al., 2025 \cite{dramko2025idioms} & \textbf{Performance} & Decompiled Code & Source Code & Fine-Tuning & \makecell[c]{AMS \\ GTV} & FTS & H & H & M \\ \hline

\rowcolor[HTML]{FFFFFF}
15 & Faruque et al., 2025 \cite{faruque2025gate} & \textbf{Performance} & Other (netlist) & Source Code & Prompting & \makecell[c]{AMS \\ GTV} & PC/FS & H & M & H \\ \hline

\rowcolor[HTML]{F6F8F9}
16 & Greiner et al., 2025 \cite{greiner2025generalization} & \makecell[c]{\textbf{Performance} \\ Robustness} & Source Code & Graph & \makecell[c]{Data Generation \\ Fine-Tuning} & \makecell[c]{AMS \\ GTV} & FTS & H & M & H \\ \hline

\rowcolor[HTML]{FFFFFF}
17 & Hatahet et al., 2025 \cite{hatahet2025generating} & \makecell[c]{Discovery \\ \textbf{Interpretability} \\ Performance} & Source Code & Graph & \makecell[c]{Human Involv. \\ Prompting} & \makecell[c]{E-BA \\ GTV} & PC/FS & H & M & H \\ \hline

\rowcolor[HTML]{F6F8F9}
18 & Huang et al., 2025 \cite{huang2025malpre} & \textbf{Performance} & Decompiled Code & Natural Language & \makecell[c]{Agent-Based A. \\ Prompting} & \makecell[c]{AMS \\ GTV} & PC/FS & H & - & M \\ \hline

\rowcolor[HTML]{FFFFFF}
19 & Jiang et al., 2025 \cite{jiang2023nova} & \makecell[c]{\textbf{Performance} \\ Robustness} & Assembly Code & Source Code & \makecell[c]{Data Generation \\ Fine-Tuning} & \makecell[c]{AMS \\ GTV} & \makecell[c]{FTS \\ PT/MS} & H & H & H \\ \hline

\rowcolor[HTML]{F6F8F9}
20 & Kirla et al., 2025 \cite{kirla2025large} & \textbf{Performance} & Decompiled Code & Natural Language & \makecell[c]{Human Involv. \\ Prompting} & E-BA & PC/FS & H & M & M \\ \hline

\rowcolor[HTML]{FFFFFF}
21 & Lea et al., 2025 \cite{11391932} & \makecell[c]{\textbf{Interpretability} \\ Performance} & Assembly Code & \makecell[c]{Assembly Code \\ Natural Language} & \makecell[c]{Data Generation \\ Fine-Tuning \\ Human Involv.} & \makecell[c]{AMS \\ E-BA \\ GTV} & FTS & H & H & H \\ \hline

\rowcolor[HTML]{F6F8F9}
22 & Li \& Klabjan, 2025 \cite{li-klabjan-2025-reverse} & \makecell[c]{\textbf{Discovery} \\ Performance} & Natural Language & Natural Language & Prompting & \makecell[c]{AMS \\ E-BA \\ GTV} & PC/FS & H & - & H \\ \hline

\rowcolor[HTML]{FFFFFF}
23 & Liao et al., 2025 \cite{liao2025augmenting} & \textbf{Performance} & Decompiled Code & Decompiled Code & Prompting & \makecell[c]{AMS \\ E-BA \\ GTV} & PC/FS & H & M & M \\ \hline

\rowcolor[HTML]{F6F8F9}
24 & Liu et al., 2025 \cite{liu2025llmtaint} & \makecell[c]{Discovery \\ \textbf{Performance} \\ Robustness} & Decompiled Code & Natural Language & Prompting & \makecell[c]{AMS \\ GTV} & PC/FS & M & M & M \\ \hline

\rowcolor[HTML]{FFFFFF}
25 & Liu et al., 2025 \cite{liu2025armqwen2} & \textbf{Performance} & Assembly Code & Decompiled Code & \makecell[c]{Data Generation \\ Fine-Tuning} & \makecell[c]{AMS \\ GTV} & FTS & H & M & H \\ \hline

\rowcolor[HTML]{F6F8F9}
26 & Liu et al., 2025 \cite{liu2025function} & \makecell[c]{Discovery \\ \textbf{Interpretability} \\ Performance} & Decompiled Code & Other (function name) & Prompting & \makecell[c]{E-BA \\ GTV} & PC/FS & H & - & H \\ \hline

\rowcolor[HTML]{FFFFFF}
27 & Liu et al., 2025 \cite{liu2025llm} & \makecell[c]{Interpretability \\ \textbf{Performance}} & Assembly Code & Assembly Code & \makecell[c]{Data Generation \\ Fine-Tuning} & \makecell[c]{AMS \\ GTV} & FTS & H & M & M \\ \hline

\rowcolor[HTML]{F6F8F9}
28 & Niang et al., 2025 \cite{niang2025enhancing} & \makecell[c]{\textbf{Performance} \\ Robustness} & Source Code & Source Code & Prompting & \makecell[c]{AMS \\ GTV} & PC/FS & M & M & H \\ \hline

\rowcolor[HTML]{FFFFFF}
29 & Ouf et al., 2025 \cite{ouf2025reverse} & \makecell[c]{Discovery \\ \textbf{Interpretability} \\ Performance} & Source Code & Natural Language & \makecell[c]{Data Generation \\ Fine-Tuning \\ Human Involv. \\ Prompting} & \makecell[c]{AMS \\ GTV} & PC/FS & H & - & H \\ \hline

\rowcolor[HTML]{F6F8F9}
30 & Petrache et al., 2025 \cite{petrache2025hybrid} & \makecell[c]{\textbf{Interpretability} \\ Performance} & Assembly Code & Natural Language & \makecell[c]{Data Generation \\ Prompting} & \makecell[c]{AMS \\ GTV} & - & H & M & M \\ \hline

\rowcolor[HTML]{FFFFFF}
31 & Rukhovich et al., 2025 \cite{rukhovich2024cad} & \makecell[c]{Interpretability \\ \textbf{Performance}} & Other (3D point cloud) & Source Code & \makecell[c]{Data Generation \\ Fine-Tuning} & AMS & FTS & H & M & M \\ \hline

\rowcolor[HTML]{F6F8F9}
32 & Sha et al., 2025 \cite{sha2025llasm} & \textbf{Performance} & Assembly Code & Other (function name) & \makecell[c]{Data Generation \\ Fine-Tuning} & AMS & FTS & H & M & H \\ \hline

\rowcolor[HTML]{FFFFFF}
33 & Siala et al., 2025 \cite{siala2025towards} & \makecell[c]{\textbf{Performance} \\ Robustness} & Source Code & Other (OCL specification) & \makecell[c]{Data Generation \\ Fine-Tuning} & \makecell[c]{AMS \\ E-BA} & FTS & H & M & M \\ \hline

\rowcolor[HTML]{F6F8F9}
34 & Su et al., 2025 \cite{su2025disco} & \makecell[c]{Interpretability \\ \textbf{Performance}} & Graph & Source Code & Prompting & \makecell[c]{AMS \\ E-BA \\ GTV} & PC/FS & H & - & H \\ \hline

\rowcolor[HTML]{FFFFFF}
35 & Villa et al., 2025 \cite{ExposingtheGuardrails} & \makecell[c]{Discovery \\ \textbf{Robustness}} & Natural Language & \makecell[c]{Natural Language \\ Other (image)} & \makecell[c]{Data Generation \\ Prompting} & AMS & PC/FS & - & M & H \\ \hline

\rowcolor[HTML]{F6F8F9}
36 & Wei et al., 2025 \cite{wei2025exposing} & \makecell[c]{Discovery \\ \textbf{Robustness}} & Natural Language & Natural Language & \makecell[c]{Human Involv. \\ Prompting} & \makecell[c]{AMS \\ GTV} & PC/FS & M & - & M \\ \hline

\rowcolor[HTML]{FFFFFF}
37 & Wong et al., 2025 \cite{wong2025decllm} & \textbf{Performance} & Decompiled Code & Source Code & Prompting & \makecell[c]{AMS \\ GTV} & PC/FS & M & M & H \\ \hline

\rowcolor[HTML]{F6F8F9}
38 & Yao et al., 2025 \cite{yao2025hdldebugger} & \makecell[c]{Interpretability \\ \textbf{Performance}} & Source Code & Source Code & \makecell[c]{Data Generation \\ Fine-Tuning \\ RAG} & \makecell[c]{AMS \\ E-BA \\ GTV} & FTS & H & M & M \\ \hline

\rowcolor[HTML]{FFFFFF}
39 & Zhang et al., 2025 \cite{zhang2025rebert} & \makecell[c]{\textbf{Performance} \\ Robustness} & Other (gate-level netlist) & Other (word-level groupings) & \makecell[c]{Data Generation \\ Fine-Tuning} & \makecell[c]{AMS \\ GTV} & FTS & H & M & M \\ \hline

\rowcolor[HTML]{F6F8F9}
40 & Asmita et al., 2024\cite{oliinyk2024fuzzing} & \makecell[c]{Discovery \\ \textbf{Performance}} & \makecell[c]{Natural Language \\ Other (Black-box binaries)} & Other (BusyBox awk applet commands) & \makecell[c]{Data Generation \\ Prompting} & \makecell[c]{AMS \\ GTV} & PC/FS & M & M & M \\ \hline

\rowcolor[HTML]{FFFFFF}
41 & Duan et al., 2024 \cite{duan2024sbcm} & \makecell[c]{Interpretability \\ \textbf{Performance}} & Graph & Natural Language & Prompting & \makecell[c]{AMS \\ GTV} & PC/FS & H & - & M \\ \hline

\rowcolor[HTML]{F6F8F9}
42 & Hu et al., 2024 \cite{hu2024degpt} & \makecell[c]{Interpretability \\ \textbf{Performance}} & Decompiled Code & Decompiled Code & \makecell[c]{Data Generation \\ Prompting} & \makecell[c]{AMS \\ E-BA} & PC/FS & H & H & H \\ \hline

\rowcolor[HTML]{FFFFFF}
43 & Ma et al., 2024 \cite{ma2024one} & \makecell[c]{\textbf{Discovery} \\ Performance} & Natural Language & Graph & \makecell[c]{Data Generation \\ Prompting} & \makecell[c]{AMS \\ GTV} & PC/FS & - & M & M \\ \hline

\rowcolor[HTML]{F6F8F9}
44 & She et al., 2024 \cite{she2024wadec} & \makecell[c]{Interpretability \\ \textbf{Performance} \\ Robustness} & Assembly Code & Source Code & \makecell[c]{Data Generation \\ Fine-Tuning} & \makecell[c]{AMS \\ E-BA \\ GTV} & FTS & H & H & H \\ \hline

\rowcolor[HTML]{FFFFFF}
45 & Tan et al., 2024 \cite{tan2024llm4decompile} & \textbf{Performance} & Assembly Code & Source Code & \makecell[c]{Data Generation \\ Fine-Tuning} & \makecell[c]{AMS \\ GTV} & FTS & H & M & H \\ \hline

\rowcolor[HTML]{F6F8F9}
46 & Walton et al., 2024 \cite{walton2024exploring} & \makecell[c]{Interpretability \\ \textbf{Performance}} & Decompiled Code & Natural Language & Prompting & \makecell[c]{AMS \\ GTV} & PC/FS & H & H & M \\ \hline

\rowcolor[HTML]{FFFFFF}
47 & Xie et al., 2024 \cite{xie2024resym} & \makecell[c]{Interpretability \\ \textbf{Performance}} & Decompiled Code & \makecell[c]{Source Code \\ Natural Language} & \makecell[c]{Data Generation \\ Fine-Tuning} & \makecell[c]{AMS \\ GTV} & FTS & H & H & M \\ \hline

\rowcolor[HTML]{F6F8F9}
48 & Zhou et al., 2024 \cite{zhou2024enhancing} & \makecell[c]{\textbf{Discovery} \\ Robustness} & Assembly Code & Source Code & \makecell[c]{Human Involv. \\ Prompting} & - & PC/FS & - & - & - \\ \hline

\end{longtable}
\endgroup

\subsection{Quantitative Landscape}
\label{quantitative_landscape}
Based on the corpus construction procedure described in Section~\ref{sec:corpus_selection_methodology}, the six-dimensional taxonomy introduced in Section~\ref{sec:taxonomy}, and the study-level classification summarized in Table~\ref{tab:llm-re-classification}, this subsection provides a quantitative overview of the LLM-based RE literature. 
The analysis characterizes both the temporal evolution of the field and the distribution of the 48 peer-reviewed studies across the taxonomy dimensions. 
Since a single study may involve multiple motivations, representations, methods, evaluation strategies, or scales, attributes within each taxonomy dimension are coded using a multi-label scheme unless explicitly stated otherwise.

\begin{figure*}[b]
\centering

\begin{minipage}[t]{0.497\linewidth}
    \centering
    \includegraphics[width=\linewidth]{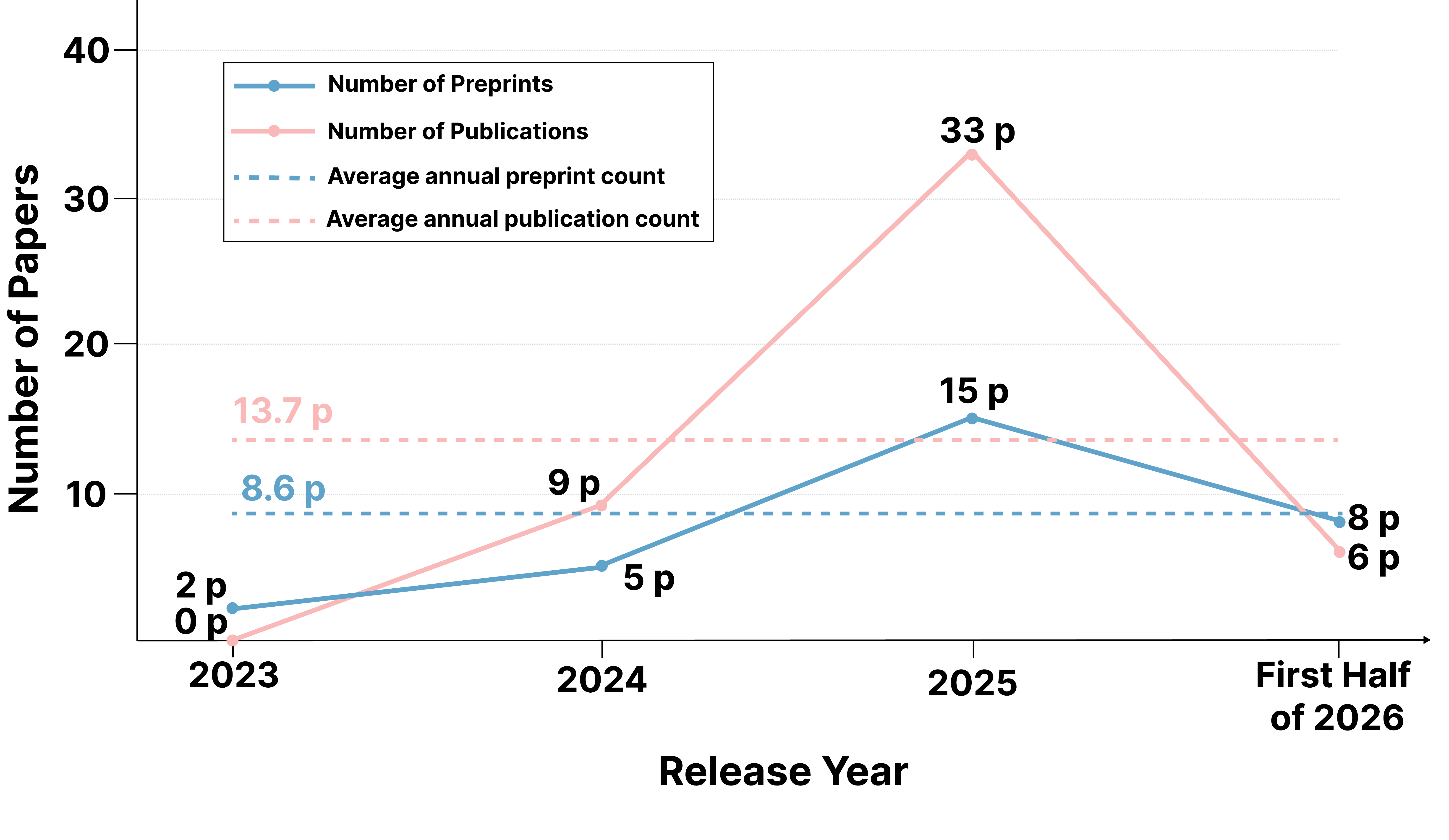}
    \caption{Annual counts and trends of published papers and preprints on LLM-based RE from 2023 to 2026. The 2026 count reflects papers identified in the first half of the year. P denotes "paper(s)".}
    \label{fig:papers_year}
\end{minipage}
\hfill
\begin{minipage}[t]{0.497\linewidth}
    \centering
    \includegraphics[width=\linewidth]{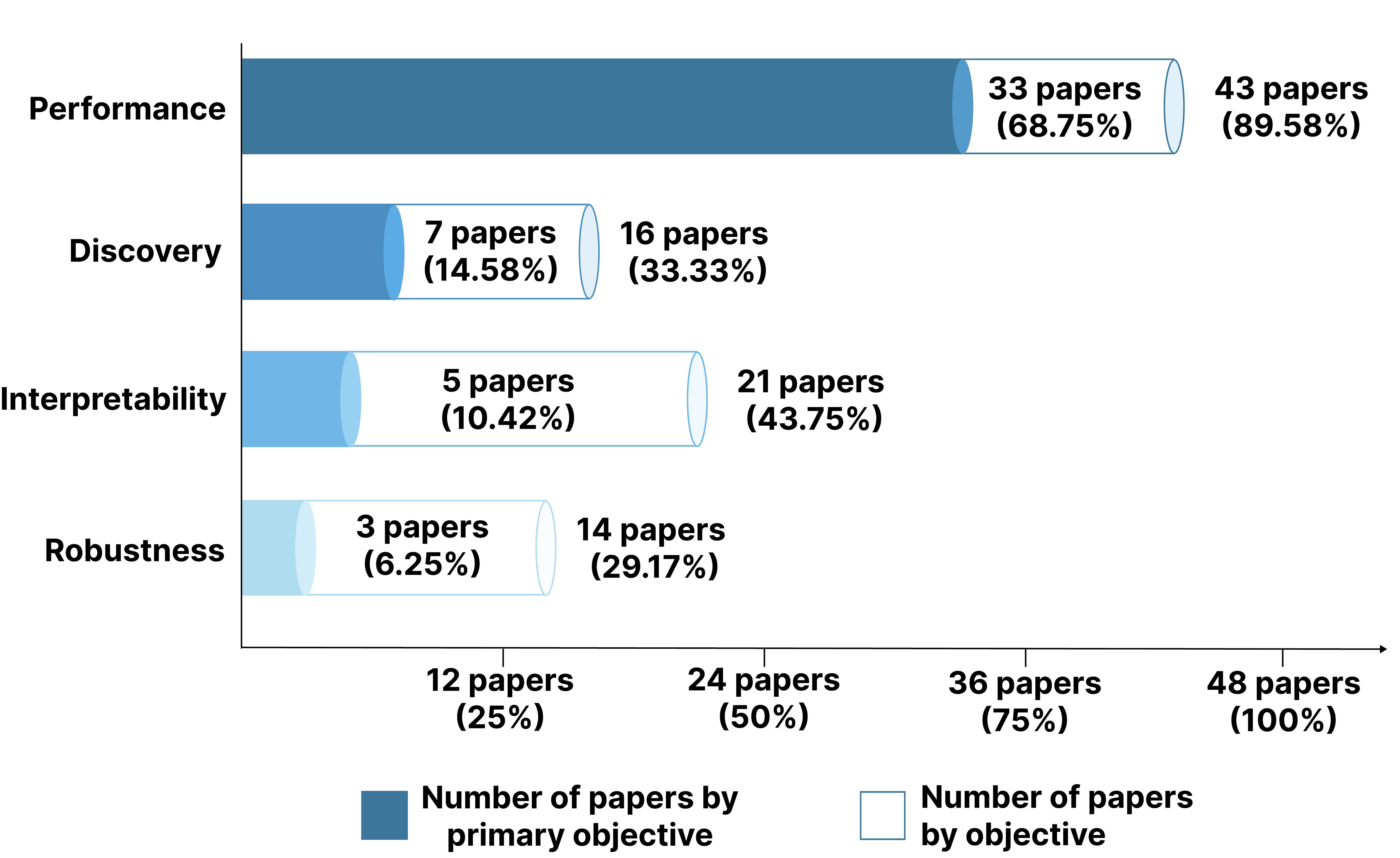}
    \caption{Distribution of 48 published LLM-based RE studies by objective. Blue bars show primary counts and white show additional non-primary counts. Percentages are counts divided by 48.}
    \label{fig:primaryobjective}
\end{minipage}
\end{figure*}

Figure~\ref{fig:papers_year} shows the temporal evolution of LLM-based RE research as of July 1, 2026. 
The first preprint appeared in 2023, followed by the first peer-reviewed publication in 2024. 
To reflect emerging research attention, this figure includes both peer-reviewed publications and preprints. 
However, preprints are excluded from the taxonomy-based analysis to maintain a stable and verifiable evidence base. 
The 2026 count should not be interpreted as a full-year estimate, as the observation window ends on July 1 and may omit studies that are still under review, awaiting formal publication, or not yet indexed. 
Overall, the trend indicates a rapid expansion of LLM-based RE, from $2$ identified studies in 2023, to $14$ in 2024, and $48$ in 2025, with $14$ additional studies already observed in the partial 2026 window. 
This pattern suggests that the field remains in an early but active stage. 

To analyze the motivations behind existing studies, we classify objectives from two perspectives: a non-exclusive view that records all objectives addressed by a study, and a primary objective view that identifies the dominant objective of each study. 
As shown in Figure~\ref{fig:primaryobjective}, performance-oriented research constitutes the largest group under both coding schemes. 
In the non-exclusive view, 89.58\% of the 48 studies consider performance-related goals, while nearly three quarters of the studies identify performance as their primary objective. 
This distribution indicates that improving effectiveness, efficiency, or task accuracy remains the central motivation in current LLM-based RE research.
At the same time, interpretability has become a substantial secondary focus. 
As LLM-based RE is situated within cybersecurity, many studies increasingly recognize the risks introduced by opaque black-box or gray-box models, especially when their outputs are used to support security-critical decisions~\cite{li-klabjan-2025-reverse, ExposingtheGuardrails, wei2025exposing}. 
Interpretability-oriented studies therefore examine whether LLM-based methods can produce understandable and faithful explanations in specific RE task settings. 
Nearly half of the surveyed studies address interpretability in some form, suggesting that the field is beginning to move beyond performance optimization toward questions of transparency, trustworthiness, and human-centered analysis. 
By contrast, Robustness and Discovery appear less frequently, indicating that systematic stress testing and exploratory use of LLMs for uncovering overlooked RE problems remain comparatively underdeveloped. 

Table~\ref{tab:summary-three-tables} summarizes the distribution of the representation, operation, validation, magnitude, and reliability axes. 
The representation axis results show that existing studies are concentrated around decompiled code as input and natural language as output, whereas lower-level targets, such as raw bytes, and structurally richer targets, such as graphs, remain less frequently explored. 
This pattern suggests that current LLM-based RE studies often favor representations that are more compatible with language-model reasoning, rather than directly modeling raw machine-level artifacts or complex relational structures.
Methodologically, prompting is the most common operational strategy, followed by data generation. 
This distribution indicates that current LLM-based RE research relies heavily on accessible uses of general-purpose LLMs, either as direct task solvers through prompting or as auxiliary tools for constructing and augmenting RE data. 
The prominence of data generation also reflects the persistent scarcity of labeled RE datasets, whereas fine-tuning, retrieval-augmented generation, and agent-based analysis remain less common, suggesting that more specialized and workflow-aware RE systems are still developing. 
In terms of validation, automated metric scoring and ground-truth validation are the dominant strategies, appearing in 44 and 39 studies, respectively, whereas expert-based assessment is used in only 17 studies. 
These results suggest that current evaluations are largely driven by scalable quantitative metrics and reference-based checks, while human-centered assessment of semantic adequacy, interpretability, and analyst usability remains less consistently incorporated. 
Consequently, reported results may still be difficult to compare when studies adopt different metrics, ground-truth definitions, and levels of expert involvement. 
The magnitude axis results further show that most studies operate at proof-of-concept/few-shot scale, with comparatively fewer efforts involving large-scale model adaptation or pretraining. 
Complementing this observation, the reliability-axis results indicate that dataset source reliability, preprocessing transparency, and ground-truth validation reliability are not uniformly reported across the literature. 
Thus, even when studies report promising results, the empirical basis for reproducing data construction, preprocessing pipelines, and label validation often remains incomplete. 
Overall, these distributions indicate a field that is methodologically diverse and rapidly expanding, but still uneven in evaluation maturity, training scale, and reproducible empirical support.

\begin{table*}[b]
\centering
\caption{Distribution of 48 published studies by target, method, evaluation, training scale, and data quality, corresponding to the representation, operation, validation, magnitude, and reliability axes, respectively.}
\label{tab:summary-three-tables}

\scriptsize
\setlength{\tabcolsep}{2pt}
\renewcommand{\arraystretch}{1.15}

\begin{minipage}[t]{0.34\textwidth}
\vspace{0pt}
\centering

\makebox[\linewidth][l]{}

\vspace{4pt}

{\renewcommand{\arraystretch}{1.316}
\resizebox{\linewidth}{!}{%
\begin{tabular}{%
>{\centering\arraybackslash}m{0.30\linewidth}|
>{\centering\arraybackslash}m{0.32\linewidth}|
>{\centering\arraybackslash}m{0.14\linewidth}|
>{\centering\arraybackslash}m{0.14\linewidth}}
\toprule
\toprule
\textbf{Dimension} & \textbf{Attribute} & \textbf{End} & \textbf{Total} \\
\midrule

\multirowcell{14}{\textbf{Target}}
& \multirowcell{2}{Assembly Code}
& Input & 10 \\
& & Output & 3 \\
\cline{2-4}

& \multirowcell{2}{Decompiled Code}
& Input & 16 \\
& & Output & 3 \\
\cline{2-4}

& \multirowcell{2}{Graph}
& Input & 3 \\
& & Output & 5 \\
\cline{2-4}

& \multirowcell{2}{Natural Language}
& Input & 6 \\
& & Output & 17 \\
\cline{2-4}

& \multirowcell{2}{Raw Bytes}
& Input & 1 \\
& & Output & -- \\
\cline{2-4}

& \multirowcell{2}{Source Code}
& Input & 10 \\
& & Output & 15 \\
\cline{2-4}

& \multirowcell{2}{Other}
& Input & 4 \\
& & Output & 10 \\

\bottomrule
\bottomrule
\end{tabular}%
}}
\end{minipage}
\hfill
\begin{minipage}[t]{0.31\textwidth}
\vspace{0pt}
\centering

\makebox[\linewidth][l]{}

\vspace{4pt}

{\renewcommand{\arraystretch}{1.45}
\resizebox{\linewidth}{!}{%
\begin{tabular}{%
>{\centering\arraybackslash}m{0.30\linewidth}|
>{\centering\arraybackslash}m{0.46\linewidth}|
>{\centering\arraybackslash}m{0.18\linewidth}}
\toprule
\toprule
\textbf{Dimension} & \textbf{Attribute} & \textbf{Total} \\
\midrule

\multirowcell{6}{\textbf{Method}}
& Agent-Based Analysis & 3 \\
\cline{2-3}
& Data Generation & 24 \\
\cline{2-3}
& Fine-Tuning & 19 \\
\cline{2-3}
& Human Involvement & 10 \\
\cline{2-3}
& Prompting & 32 \\
\cline{2-3}
& RAG & 3 \\
\midrule

\multirowcell{3}{\textbf{Evaluation}}
& AMS & 44 \\
\cline{2-3}
& E-BA & 17 \\
\cline{2-3}
& GTV & 39 \\
\midrule

\multirowcell{3}{\textbf{Training Scale}}
& PC/FS & 29 \\
\cline{2-3}
& FTS & 18 \\
\cline{2-3}
& PT/MS & 1 \\

\bottomrule
\bottomrule
\end{tabular}%
}}
\end{minipage}
\hfill
\begin{minipage}[t]{0.31\textwidth}
\vspace{0pt}
\centering

\makebox[\linewidth][l]{}

\vspace{4pt}

\resizebox{\linewidth}{!}{%
\begin{tabular}{%
>{\centering\arraybackslash}m{0.32\linewidth}|
>{\centering\arraybackslash}m{0.32\linewidth}|
>{\centering\arraybackslash}m{0.16\linewidth}|
>{\centering\arraybackslash}m{0.16\linewidth}}
\toprule
\toprule
\textbf{Dimension} & \textbf{Attribute} & \textbf{Level} & \textbf{Total} \\
\midrule

\multirowcell{9}{\textbf{Data Quality}}
& \multirowcell{3}{GTV\\Reliability}
& High & 36 \\
& & Medium & 9 \\
& & -- & 3 \\
\cline{2-4}

& \multirowcell{3}{Preprocessing\\Assumption}
& High & 8 \\
& & Medium & 29 \\
& & -- & 11 \\
\cline{2-4}

& \multirowcell{3}{Source\\Reliability}
& High & 23 \\
& & Medium & 24 \\
& & -- & 1 \\

\bottomrule
\bottomrule
\end{tabular}%
}
\end{minipage}

\end{table*}

\begin{figure*}[!htbp]
    \centering

    \captionsetup[subfigure]{
        position=top,
        labelformat=parens,
        labelsep=none,
        justification=raggedright,
        singlelinecheck=false,
        font=footnotesize,
        labelfont=bf,
        skip=4pt
    }
    \begin{subfigure}{\textwidth}
        \label{fig:opensourcecomparison}
        \includegraphics[width=\linewidth]{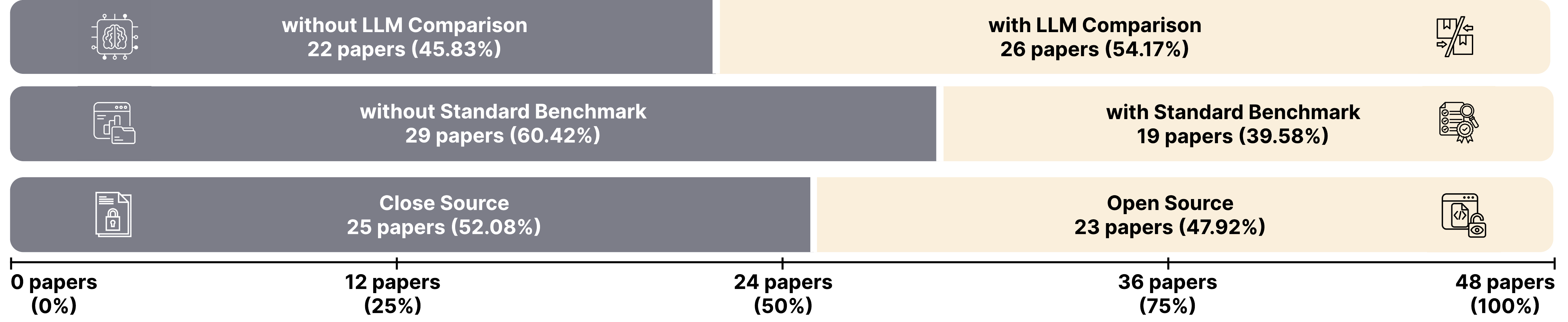}
    \end{subfigure}
    
    \caption{Distribution of 48 published studies by open/closed-source availability, use of standard benchmarks, and LLMs for comparison.}
    \label{fig:study_comparisons}
\end{figure*}

Table~\ref{tab:implementation-evaluation-settings} further details the research configurations reported by the collected studies, including open-source status, task definitions, LLM choices, detailed representations, benchmarks, and hardware/software. 
This table shows that LLM-based RE studies are highly heterogeneous not only in their target tasks, but also in the models, representations, benchmarks, and implementation environments used for evaluation. 
Figure~\ref{fig:study_comparisons} summarizes three reproducibility-related indicators derived from these configurations. 
Only 23 studies (47.92\%) provide open-source resources, while 25 studies (52.08\%) do not, which further constrains independent verification and reuse. 
Similarly, 19 studies (39.58\%) use standard benchmarks, while 29 studies (60.42\%) rely on self-constructed or non-standard evaluation settings, indicating that benchmark standardization remains limited. 
Finally, only 26 studies (54.17\%) include LLM comparisons, whereas 22 studies (45.83\%) do not, suggesting that many evaluations still lack systematic comparison across model choices. 
These results show that although LLM-based RE research is becoming increasingly diverse in tasks and technical configurations, its empirical foundation remains uneven: standard benchmarks, comparative model evaluation, and open-source artifacts are not yet consistently adopted across the field.


\begingroup
\scriptsize
\setlength{\tabcolsep}{1pt}
\renewcommand{\arraystretch}{1.05}
\setlength\LTleft{0pt}
\setlength\LTright{0pt}
\emergencystretch=1em
\hyphenpenalty=10000
\exhyphenpenalty=10000
\newcommand{\wrapcell}[2]{%
  \begin{minipage}[t]{#1}
  \raggedright
  \hbadness=10000
  \hfuzz=2pt
  #2
  \end{minipage}%
}

\begin{longtable}{|
>{\centering\arraybackslash}p{0.015\textwidth}|
>{\raggedright\arraybackslash}p{0.095\textwidth}|
>{\centering\arraybackslash}p{0.015\textwidth}|
>{\raggedright\arraybackslash}p{0.110\textwidth}|
>{\raggedright\arraybackslash}p{0.155\textwidth}|
>{\raggedright\arraybackslash}p{0.135\textwidth}|
>{\raggedright\arraybackslash}p{0.16\textwidth}|
>{\raggedright\arraybackslash}p{0.26\textwidth}|
}

\caption{The settings of collected 48 published studies on LLM-based RE, including author-year references with corresponding numbered in-text citations, open-source status (OS), task, and configurations. The setting category includes subcolumns for LLM, representation, benchmark, and conditions. In the LLM subcolumn, an * indicates an LLM used for comparison; in the Benchmark subcolumn, an * indicates a standard benchmark reported by the authors. In the Configurations subcolumn, HW denotes hardware and SW denotes software; all abbreviations listed under Configurations are taken from the corresponding original papers.}
\label{tab:implementation-evaluation-settings} \\

\hline
\rowcolor[HTML]{F6F8F9}
\multicolumn{1}{|c|}{\textbf{ID}} &
\multicolumn{1}{c|}{\textbf{Reference}} &
\multicolumn{1}{c|}{\textbf{OS}} &
\multicolumn{1}{c|}{\textbf{Task}} &
\multicolumn{4}{c|}{\textbf{Settings}} \\ 
\hline

\rowcolor[HTML]{F6F8F9}
& & & &
\multicolumn{1}{c|}{\textbf{LLM}} &
\multicolumn{1}{c|}{\textbf{Representation}} &
\multicolumn{1}{c|}{\textbf{Benchmark}} &
\multicolumn{1}{c|}{\textbf{Configurations}} \\ 
\hline
\endfirsthead

\hline
\rowcolor[HTML]{F6F8F9}
\multicolumn{1}{|c|}{\textbf{ID}} &
\multicolumn{1}{c|}{\textbf{Reference}} &
\multicolumn{1}{c|}{\textbf{OS}} &
\multicolumn{1}{c|}{\textbf{Task}} &
\multicolumn{4}{c|}{\textbf{Settings}} \\ 
\hline

\rowcolor[HTML]{F6F8F9}
& & & &
\multicolumn{1}{c|}{\textbf{LLM}} &
\multicolumn{1}{c|}{\textbf{Representation}} &
\multicolumn{1}{c|}{\textbf{Benchmark}} &
\multicolumn{1}{c|}{\textbf{Configurations}} \\ 
\hline
\endhead

\rowcolor[HTML]{FFFFFF}
1 & \conditions{Achamyeleh \\et al., 2026 \cite{achamyeleh2026helios}} & N & Structure-aware binary decompilation & \wrapcell{0.115\textwidth}{*Gemini-2.0-Flash \\ *GPT-4.1 Mini} & Decompiled pseudo C code and graph information & \wrapcell{0.115\textwidth}{*HumanEval-Decompile \\ *MBPP-Decompile} & \conditions{\textbf{HW:} One NVIDIA A100 GPU \\ \textbf{SW:} GCC 11.4 toolchain; Ghidra 11.0; Python 3.10} \\ \hline

\rowcolor[HTML]{F6F8F9}
2 & \conditions{Basque et al., \\2026 \cite{decompiling}} & Y & Human–LLM teaming in software RE & \wrapcell{0.145\textwidth}{GPT-4o (default) \\ *Claude-3-5-Sonnet \\ *Gemini-Pro \\ *GPT-4-Turbo \\ *GPT-4o-Mini\rule[-1.2ex]{0pt}{0pt}} & IDA-based decompiled binary view with prompt-driven annotation features & \wrapcell{0.115\textwidth}{153-person software RE online survey 2\\ Capture-the-Flag-style challenges} & \conditions{\textbf{HW:} Virtual machine \\ \textbf{SW:} Custom decompiler plugin; DOJO; IDA Pro 9} \\ \hline

\rowcolor[HTML]{FFFFFF}
3 & \conditions{Sha et al., \\2026 \cite{sha2025hyres}} & Y & Stripped binary data structure recovery & Qwen-2 72B & Normalized decompiled C code, LLVM IR & \wrapcell{0.115\textwidth}{OSPERY \\ *BinaryCorp} & \conditions{\textbf{HW:} 96 CPU cores with 1 TB RAM; 128 CPU cores with 2 TB RAM \\ \textbf{SW:} Datalog; GTE model; IDA Pro; LLVM IR} \\ \hline

\rowcolor[HTML]{F6F8F9}
4 & \conditions{Surk et al., \\2026 \cite{app16062801}} & Y & Requirements recovery from source code & Gemini-1.5-pro & \wrapcell{0.130\textwidth}{Preprocessed source code \& product manager request in a \\long-context cached prompt} & WordPress, ERPNext, Ghost, and Odoo (Case Study) & \conditions{\textbf{HW:} - \\ \textbf{SW:} Google Cloud Vertex AI; Python 3.11.1; 2M-token context window} \\ \hline

\rowcolor[HTML]{FFFFFF}
5 & \conditions{Xue et al., \\2026 \cite{xue2026record}} & N & RE codebase to causal relational diagram & \wrapcell{0.115\textwidth}{Claude 3.7 Sonnet \\ *Phi-4-RFT \\ *Qwen2.5-7/14B-Instruct-RFT \\ *Qwen3-32B\rule[-1.2ex]{0pt}{0pt}} & Source code and document derived causal graph & \wrapcell{0.115\textwidth}{*News Vendor \\ *MiniSCOT \\ *Black-Scholes} & \conditions{\textbf{HW:} AWS g5.48xlarge (Eight NVIDIA A10G GPUs; 192 vCPUs; 768 GB RAM); AWS p3.16xlarge (Eight NVIDIA Tesla V100 GPUs; 64 vCPUs; 488 GB RAM) \\ \textbf{SW:} Abstract syntax tree (AST); data flow analysis} \\ \hline

\rowcolor[HTML]{F6F8F9}
6 & \conditions{Yang et al., \\2026 \cite{YANG2026104468}} & N & Industrial control proprietary protocol RE & \wrapcell{0.115\textwidth}{DeepSeek-R1-0528 \\ *DeepSeek- R1 \\ *Qwen3-235B \\ *Qwen3-235B-thinking\rule[-1.2ex]{0pt}{0pt}} & \wrapcell{0.115\textwidth}{Raw protocol data with fused multi-source \\ protocol knowledge} & Industrial control proprietary protocol benchmark & \conditions{\textbf{HW:} Intel i7-10700 CPU; 64 GB RAM \\ \textbf{SW:} Boofuzz; Chroma vector database; Tshark/Wireshark; Ubuntu 20.04} \\ \hline

\rowcolor[HTML]{FFFFFF}
7 & \conditions{Alazraki et al., \\2025 \cite{alazraki2026reverse}} & N & Human preference RE in judge-LLMs & \wrapcell{0.120\textwidth}{*Claude Haiku \\ *Command R (35B) \\ *Command R+ (104B) \\ *Command R7B \\ *GPT-3.5 \\ *GPT-4o-mini \\ *Llama 3.1 8/70B Instruct\rule[-1.2ex]{0pt}{0pt}} & Natural language text from preambles and questions & \wrapcell{0.115\textwidth}{*MT-Bench \\ *Arena-Hard benchmark \\ *UltraFeedback} & \conditions{\textbf{HW:} Google Cloud TPU v5e (64 chips); One NVIDIA H100 GPU \\ \textbf{SW:} Command} \\ \hline

\rowcolor[HTML]{F6F8F9}
8 & \conditions{Arai et al., \\2025 \cite{arai2025staged}} & N & Original Entry Point (OEP) identification in packed binaries & \wrapcell{0.115\textwidth}{*Llama2-7B \\ *Llama3.1-8B \\ *Llama3.2-3B \\ *Qwen2.5-3/7B \\ *Qwen2.5-Coder-3B/7B\rule[-1.2ex]{0pt}{0pt}} & Assembly code from binaries & Binary dataset from Github public datasets UPX, ASPack and PECompact & \conditions{\textbf{HW:} Intel Core i5-12400F CPU; Two NVIDIA GeForce RTX 4060 Ti (16 GB); Two 32 GB DDR4 RAM \\ \textbf{SW:} llama.cpp b5943; LLaMA-Factory 0.9.3.dev0; Ollama 0.9.6; Ubuntu 24.04 LTS} \\ \hline

\rowcolor[HTML]{FFFFFF}
9 & \conditions{Arasteh et al., \\2025 \cite{arasteh2025trim}} & N & Functionality retrieval in cyber-physical system binaries & \wrapcell{0.115\textwidth}{*CodeQwen \\ *CodeStral \\ *DeepSeek-Coder\rule[-1.2ex]{0pt}{0pt}} & Decompiled firmware functions grouped into modules & ArduPilot FMUV3 benchmark (4 devices; 467/476 modules inconsistently) & \conditions{\textbf{HW:} CentOS-based server; Two GPUs; 48 GB RAM \\ \textbf{SW:} angr; Ghidra; lm-eval-harness; Python} \\ \hline

\rowcolor[HTML]{F6F8F9}
10 & \conditions{Boronat et al., \\2025 \cite{boronat2025mdre}} & Y & Domain models recovery from Java codebases & \wrapcell{0.115\textwidth}{*gpt-4omini-2024-07-18 \\ *gemini-1.5-flash} & Java source code & ModelSet (*standard and custom) & \conditions{\textbf{HW:} - \\ \textbf{SW:} Legacy/large-scale systems; Twelve Java projects (JavaFX/API/Android) with three granularity} \\ \hline

\rowcolor[HTML]{FFFFFF}
11 & \conditions{Brown et al., \\2025 \cite{brown2025large}} & N & LLM-Driven immersive visualization for binary RE & GPT-4.1 and o4-mini & Binary-analysis tool outputs and decompiled pseudo-source & \wrapcell{0.150\textwidth}{hexdump (pilot study) \\ V11 (pilot study, bot simulation)} & \conditions{\textbf{HW:} VR environment with tracked head-mounted display and hand controllers \\ \textbf{SW:} angr; Capa; Ghidra; MCP Oxide server; MCP visualization server; Oxide Framework; Python; Unity} \\ \hline

\rowcolor[HTML]{F6F8F9}
12 & \conditions{Chen et al., \\2025 \cite{chen2025suigpt}} & Y & LLM-assisted move smart contract decompilation & GPT-4o/GPT-4 & Decompiled Sui Move bytecode & \wrapcell{0.115\textwidth}{*10 official Sui Move unit-test examples \\ 30 real-world Sui contracts \\ 12 user studies\rule[-1.2ex]{0pt}{0pt}} & \conditions{\textbf{HW:} - \\ \textbf{SW:} Sui blockchain; Revela Decompiler; Move Disassembler} \\ \hline

\rowcolor[HTML]{FFFFFF}
13 & \conditions{Choi et al., \\2025 \cite{choi2024chatdeob}} & N & Obfuscated source code deobfuscation & GPT-3.5-turbo-0125 & C/C++ source code & \wrapcell{0.120\textwidth}{*Juliet C/C++ 1.3 \\ Obfuscation Benchmarks\rule[-1.2ex]{0pt}{0pt}} & \conditions{\textbf{HW:} - \\ \textbf{SW:} C; Tigress; LLVM; Linux} \\ \hline

\rowcolor[HTML]{F6F8F9}
14 & \conditions{Dramko et al., \\2025 \cite{dramko2025idioms}} & Y & LLM-based neural decompilation with definition generation & \wrapcell{0.115\textwidth}{*CodeGemma-2b/7b \\ *CodeLlama-7b \\ *CodeQwen2.5-0.5b \\ *LLM4Decompile-1.3b-v2\rule[-1.2ex]{0pt}{0pt}} & Decompiled pseudo code from C & \wrapcell{0.115\textwidth}{*EXEBENCH \\ REALTYPE} & \conditions{\textbf{HW:} - \\ \textbf{SW:} C; Hex-Rays; Linux; GCC; GHCC; Ghidra; Python} \\ \hline

\rowcolor[HTML]{FFFFFF}
15 & \conditions{Faruque et al., \\2025 \cite{faruque2025gate}} & N & Register Transfer Level (RTL) recovery from gate-level netlist & \wrapcell{0.115\textwidth}{*Gemini-2.5-Flash \\ *Gemini-2.5-Flash-Thinking \\ *GPT-4.1-Mini \\ *O3-Mini\rule[-1.2ex]{0pt}{0pt}} & netlist N from RTL & *VerilogEval & \conditions{\textbf{HW:} - \\ \textbf{SW:} ABC tool within Yosys; Icarus Verilog; Python} \\ \hline

\rowcolor[HTML]{F6F8F9}
16 & \conditions{Greiner et al., \\2025 \cite{greiner2025generalization}} & Y & UML sequence diagram recovery from source code & Code-t5-small & Java/Python method code paired with JSON textual UML sequence diagrams & Datasets from CodeSearchNet & \conditions{\textbf{HW:} - \\ \textbf{SW:} Java; JavaParser; Python; Python’s built-in AST module} \\ \hline

\rowcolor[HTML]{FFFFFF}
17 & \conditions{Hatahet et al., \\2025 \cite{hatahet2025generating}} & Y & Software architecture description generation from source code & GPT-4o & \wrapcell{0.115\textwidth}{PlantUML-encoded structure, \\ C++ source code} & Coffee machine and Dishwasher from IBM Rhapsody 8.4 C++ source code (Case Study) & \conditions{\textbf{HW:} - \\ \textbf{SW:} Enterprise Architect; Python; PlantUML} \\ \hline

\rowcolor[HTML]{F6F8F9}
18 & \conditions{Huang et al., \\2025 \cite{huang2025malpre}} & Y & Malware protocol RE & Qwen2.5-72b & Code slices and decompiled code & PRE evaluation dataset & \conditions{\textbf{HW:} Intel Core i7-9750H CPU (2.60 GHz); 64 GB RAM \\ \textbf{SW:} BAAI/bge-reranker-v2-m3; Hex-Rays; IDA Pro; Joern; LangGraph; PinTool; Unicorn; Windows 10 Pro\rule[-1.2ex]{0pt}{0pt}} \\ \hline

\rowcolor[HTML]{FFFFFF}
19 & Jiang et al., 2025 \cite{jiang2023nova} & Y & Binary decompilation and similarity detection & \wrapcell{0.115\textwidth}{Nova-1.3/6.7B \\ *CodeLlama-7B \\ *DeepSeekCoder-7B \\ *GPT-3.5-Turbo \\ *GPT-4o \\ *Granite(3B-Code-128K) \\ *LLM4Decompile-1.3/6.7B \\ *LLMCompiler-7B \\ *Qwen-2.5-Coder-7B \\ *StarCoder2-7B\rule[-1.2ex]{0pt}{0pt}} & X86-64 assembly functions & \wrapcell{0.160\textwidth}{*HumanEval \\ *Binutils, Curl, ImageMagick, SQLite, OpenSSL, and Putty} & \conditions{\textbf{HW:} Eight 40GB memory NVIDIA RTX A100 GPUs \\ \textbf{SW:} GCC; Objdump; Python; PyTorch} \\ \hline

\rowcolor[HTML]{F6F8F9}
20 & Kirla et al., 2025 \cite{kirla2025large} & N &Decompiled code annotation & \wrapcell{0.115\textwidth}{Ollama(\text{qwen2.5-coder: }\\
\text{32b-instruct-q4\_K\_M})} & Ghidra Decompiled C/C++ code & 360 functions from 6 open-source programs & \conditions{\textbf{HW:} VRAM of a 24 GB consumer GPU \\ \textbf{SW:} Ghidra; Javascript} \\ \hline

\rowcolor[HTML]{FFFFFF}
21 & Lea et al., 2025 \cite{11391932} & Y & LLM-assisted x86 assembly RE with comment generation & \wrapcell{0.115\textwidth}{Qwen2.5-Coder-7B \\ *CodeGemma-7B \\ *CodeLlama-7/13/34B \\ *Qwen2.5-Coder-3/14/32B\rule[-1.2ex]{0pt}{0pt}} & Masked and completed x86 assembly code with task-specific natural language Q\&A & 5,981 x86 assembly examples (Assembly Shellcode, Rosetta Code, Shell-Storm, xorpd) and x86 assembly Q\&A pairs & \conditions{\textbf{HW:} One NVIDIA RTX 6000 GPU with 48GB of VRAM; One NVIDIA RTX 5090 \\ \textbf{SW:} IDA-64; OpenAI Triton kernels; Unsloth framework} \\ \hline

\rowcolor[HTML]{F6F8F9}
22 & Li \& Klabjan, 2025 \cite{li-klabjan-2025-reverse} & Y & Black-box prompt recovery & \wrapcell{0.115\textwidth}{GPT-3.5 \\ *Llama-2 Chat (7B)} & Five-output response sets for prompt reconstruction & \wrapcell{0.115\textwidth}{REprompt \\ REhard \\ REeasy} & \conditions{\textbf{HW:} - \\ \textbf{SW:} Git} \\ \hline

\rowcolor[HTML]{FFFFFF}
23 & Liao et al., 2025 \cite{liao2025augmenting} & Y & Function boundary, variable type, and contract attribute recovery & \wrapcell{0.120\textwidth}{ChatGPT (GPT-3.5) \\ *Deepseek-v3 671B \\ *GPT-4o mini \\ *Llama-3 7B \\ *Qwen-2.5-coder 32B} & Decompiler output from Ethereum Virtual Machine bytecode compiled from Solidity code, with instructions as prompts for LLMs & 456 decompiler method pairs & \conditions{\textbf{HW:} Intel i9-10980XE CPU; NVIDIA RTX 3090 GPU; 250 GB RAM \\ \textbf{SW:} Dedaub/Gigahorse; Python 3.8.10; SmartState; Solidity compiler; Tree-sitter; Ubuntu 20.04 server; Z3 solver} \\ \hline

\rowcolor[HTML]{F6F8F9}
24 & Liu et al., 2025 \cite{liu2025llmtaint} & N & Binary vulnerability inspection via taint analysis & \wrapcell{0.115\textwidth}{GPT-4.0 \\ *GPT-3.5} & Decompiled pseudocode from compiled C/C++ binaries & \wrapcell{0.160\textwidth}{*Juliet Test Suite v1.3 \\ *Karonte firmware} & \conditions{\textbf{HW:} - \\ \textbf{SW:} Ghidra; Python} \\ \hline

\rowcolor[HTML]{FFFFFF}
25 & Liu et al., 2025 \cite{liu2025armqwen2} & N & ARM C decompilation quality improvement & \wrapcell{0.115\textwidth}{ARMQwen2 (Qwen2 7B based) \\ *DeepSeek-Coder-6.7B \\ *GPT-4o-mini\rule[-1.2ex]{0pt}{0pt}} & Preprocessed ARM assembly code & \wrapcell{0.115\textwidth}{*HumanEval-Decompile \\ *Real-world Synth} & \conditions{\textbf{HW:} Intel Xeon Gold 5218; Two RTX A6000 48G GPUs; 256GB RAM \\ \textbf{SW:} ARM GCC; ARM objdump; LLaMA-Factory; Python; Ubuntu 22.04.4 LTS} \\ \hline

\rowcolor[HTML]{F6F8F9}
26 & Liu et al., 2025 \cite{liu2025function} & N & Firmware function renaming & GPT-4.0 & Decompiled pseudo code & 6 firmware samples with symbol tables & \conditions{\textbf{HW:} Intel Core i7-8750H CPU; 64G RAM \\ \textbf{SW:} Ghidra; Linux; Python} \\ \hline

\rowcolor[HTML]{FFFFFF}
27 & Liu et al., 2025 \cite{liu2025llm} & N & ARM binary code deobfuscation & \wrapcell{0.115\textwidth}{Qwen-2.5 \\ *Llama3-8B \\ *Qwen2.5-7B\rule[-1.2ex]{0pt}{0pt}} & Obfuscated ARM Assembly Code & Benchmark from GNU toolkit and Juliet C/C++ 1.3, with 14 obfuscation settings & \conditions{\textbf{HW:} Five NVIDIA A6000 (48 GB) GPUs; Two Intel Xeon Gold 8352Y CPUs; 512 GB RAM \\ \textbf{SW:} Linux; LLaMA-Factory; OLLVM; Tigress} \\ \hline

\rowcolor[HTML]{F6F8F9}
28 & Niang et al., 2025 \cite{niang2025enhancing} & N & Model-driven RE-assisted code generation & LLaMA 3 & Processed Java source code & *RxJava & \conditions{\textbf{HW:} - \\ \textbf{SW:} Java; Moose platform; VerveineJ} \\ \hline

\rowcolor[HTML]{FFFFFF}
29 & Ouf et al., 2025 \cite{ouf2025reverse} & N & User story recovery from source code & \wrapcell{0.115\textwidth}{*DeepSeek-R1 \\ *GPT-4o-mini \\ *GPT-O1 \\ *LLAMA 3.1-8/70/405B\rule[-1.2ex]{0pt}{0pt}} & C++ source code snippets with prompt templates for user-story generation & IBM Project\_CodeNet\_C++1400 subset & \conditions{\textbf{HW:} - \\ \textbf{SW:} Replicate platform} \\ \hline

\rowcolor[HTML]{F6F8F9}
30 & \conditions{Petrache et \\al., 2025 \cite{petrache2025hybrid}} & N & Function name recovery in stripped binaries & DeepSeek R1 & Normalized assembly with local call-context prompts & 6,824 Windows binaries & \conditions{\textbf{HW:} Two NVIDIA RTX 4090 Ti GPUs; 20 CPU cores; 128 GiB RAM \\ \textbf{SW:} Microsoft's Vcpkg package manager; MSVC toolchain\rule[-1.2ex]{0pt}{0pt}} \\ \hline

\rowcolor[HTML]{FFFFFF}
31 & \conditions{Rukhovich \\et al., 2025 \\ \cite{rukhovich2024cad}} & Y & Translating a point cloud into Python code & \wrapcell{0.115\textwidth}{Qwen2-1.5B \\ *GPT-4o (Evaluation)} & \wrapcell{0.115\textwidth}{Sequence query tokens mapped by a lightweight projection module, \\ CadQuery Python code} & \wrapcell{0.160\textwidth}{1,000,000 CAD models in the form of sketch-extrude sequences written in Python CadQuery \\ *DeepCAD \\ *Fusion360 \\ *CC3D (2,973 models)} & \conditions{\textbf{HW:} - \\ \textbf{SW:} Point cloud projector, Python} \\ \hline

\rowcolor[HTML]{F6F8F9}
32 & Sha et al., 2025 \cite{sha2025llasm} & Y & Function naming in binaries & \wrapcell{0.115\textwidth}{llasm (llama2 based) \\ *GPT4 \\ *llama2-7b/13b} & Preserved binary program information & \wrapcell{0.115\textwidth}{*BinaryCorp \\ *Debin} & \conditions{\textbf{HW:} Eight NVIDIA GeForce RTX 4090 GPUs; Four NVIDIA Tesla A100 GPUs; 96 CPU cores with 1 TB RAM; 128 CPU cores with 2 TB RAM \\ \textbf{SW:} BPE tokenizer\rule[-1.2ex]{0pt}{0pt}} \\ \hline

\rowcolor[HTML]{FFFFFF}
33 & Siala et al., 2025 \cite{siala2025towards} & N & Object Constraint Language (OCL) specification extraction from Java and Python programs & Mistral-7B-v0.1 & Preprocessed Java and Python Source Code & Java/Python-OCL dataset (AVATAR-TC-derived) & \conditions{\textbf{HW:} - \\ \textbf{SW:} AgileUML; Eclipse OCL; Java; Python} \\ \hline

\rowcolor[HTML]{F6F8F9}
34 & Su et al., 2025 \cite{su2025disco} & Y & EVM bytecode-to-source code decompilation & \wrapcell{0.115\textwidth}{GPT-4-0315-Preview \\ *Llama3} & Semantic graph & \wrapcell{0.160\textwidth}{Dataset from Etherscan \\ Deployed contracts} & \conditions{\textbf{HW:} One Intel Xeon Gold 6252 CPU@2.10GHz; One RTX 2080 Ti GPU; 251 GB RAM \\ \textbf{SW:} Ethersolve; ILF; Solc; Ubuntu 18.04.6; Vandal} \\ \hline

\rowcolor[HTML]{FFFFFF}
35 & Villa et al., 2025 \cite{ExposingtheGuardrails} & Y & RE and jailbreaking safety filters in black-box text-to-image pipelines & GPT-4/GPT-4o & Natural-language prompts, revised prompts, rejection messages, and timing traces in a black-box T2I pipeline & Adversarial multilingual prompts & \conditions{\textbf{HW:} - \\ \textbf{SW:} Chrome developer tools; DALL·E 2; DALL·E 3} \\ \hline

\rowcolor[HTML]{F6F8F9}
36 & Wei et al., 2025 \cite{wei2025exposing} & N & Black-box RE of LLM plugins' hidden prompts & ChatGPT & Natural language text (adversarial prompts) & 5,000 deployed plugins from GPT store & \conditions{\textbf{HW:} - \\ \textbf{SW:} ChatGPT plugins; Selenium; Web crawler} \\ \hline

\rowcolor[HTML]{FFFFFF}
37 & Wong et al., 2025 \cite{wong2025decllm} & N & Recompilable decompilation & \wrapcell{0.115\textwidth}{GPT-3.5 \\ *GPT-4} & Decompiled C-like pseudocode & \wrapcell{0.160\textwidth}{*Code Contest dataset \\ *CGC binaries \\ *Coreutils 9.5 \\ *Decompile-eval \\ *CodeQL on Juliet\rule[-1.2ex]{0pt}{0pt}} & \conditions{\textbf{HW:} AMD Ryzen 3970X (32-core CPU); NVIDIA RTX A6000 GPU; 256 GB RAM \\ \textbf{SW:} AFL++; ASAN; C; GCC; Ghidra; IDA Pro; Python} \\ \hline

\rowcolor[HTML]{F6F8F9}
38 & Yao et al., 2025 \cite{yao2025hdldebugger} & N & Automatic HDL debugging and bug fixing & \wrapcell{0.115\textwidth}{CodeLlama-13B \\ *ChatGPT \\ *Deepseek \\ *GPT-4 \\ *Mistral \\ *OpenChat \\ *Orca2 \\ *StableCode \\ *StarCoder \\ *WizardCoder\rule[-1.2ex]{0pt}{0pt}} & Buggy HDL source code with error-message-aware RAG & Industry HDL code dataset & \conditions{\textbf{HW:} - \\ \textbf{SW:} HDL compiler} \\ \hline

\rowcolor[HTML]{FFFFFF}
39 & \conditions{Zhang et al., \\2025 \cite{zhang2025rebert}} & N & Word-level structure recovery from gate-level netlists & ReBERT & token sequence representing a pair of bits & *ITC'99 gate-level netlist & \conditions{\textbf{HW:} NVIDIA A100 GPU \\ \textbf{SW:} Hugging Face transformers (4.42.2); Python; PyTorch (2.3.1)} \\ \hline

\rowcolor[HTML]{F6F8F9}
40 & \conditions{Asmita et al., \\2024 \cite{oliinyk2024fuzzing}} & Y & LLM-assisted BusyBox fuzzing with crash reuse & GPT-4 & Grammar-conforming initial seeds (mainly awk scripts) for BusyBox targets & BusyBox ELF binaries & \conditions{\textbf{HW:} - \\ \textbf{SW:} AFL++; AFL-Triage; BusyBox v1.36.1; GDB; Ghidra; Python; QEMU emulator; Ubuntu 22.04} \\ \hline

\rowcolor[HTML]{FFFFFF}
41 & Duan et al., 2024 \cite{duan2024sbcm} & N & Binary code modularization & LLaMa3 & Strings, FSG(JSON) & \wrapcell{0.130\textwidth}{B1: BusyBox software \\ T1: Trusted standalone programs \\ M1: Malware collection\rule[-1.2ex]{0pt}{0pt}} & \conditions{\textbf{HW:} - \\ \textbf{SW:} Static analysis tools (They did not specify tools name they used.)} \\ \hline

\rowcolor[HTML]{F6F8F9}
42 & Hu et al., 2024 \cite{hu2024degpt} & Y & Decompiler output optimization & \wrapcell{0.115\textwidth}{gpt-3.5-turbo \\ *GPT-4} & Decompiled pseudo C code & Code from LeetCode, Coreutils, Mirai, AudioFlux & \conditions{\textbf{HW:} Intel Xeon E5-2620 v4 CPU; Two RTX 3090 GPUs; 128GB RAM; 3TB storage \\ \textbf{SW:} cinspector; GCC 9.4.0; Ghidra 10.2.3; Python; tree-sitter; Ubuntu 20.04\rule[-1.2ex]{0pt}{0pt}} \\ \hline

\rowcolor[HTML]{FFFFFF}
43 & Ma et al., 2024 \cite{ma2024one} & Y & LLM-assisted stateful fuzzing of matter IoT devices & GPT-4-Turbo & PDF-formatted Matter specification & 23 real Matter devices & \conditions{\textbf{HW:} Intel Core i7 (4.9 GHz); nRF52840 USB Dongle; 32 GB RAM \\ \textbf{SW:} chip-tool; OpenThread Border Router; ot-br-posix; Ubuntu 20.04\rule[-1.2ex]{0pt}{0pt}} \\ \hline

\rowcolor[HTML]{F6F8F9}
44 & She et al., 2024 \cite{she2024wadec} & Y & Wasm binary code decompilation & CodeLlama-7b-hf & wat code snippets (WebAssembly text format) & \wrapcell{0.160\textwidth}{Wasm with Emscripten \\ *TheAlgorithms/c \\ *bigcode/the-stack-dedup} & \conditions{\textbf{HW:} AMD EPYC 7713 (64-core CPU); Two NVIDIA A100 80GB GPUs; 516 GB RAM \\ \textbf{SW:} CUDA 12.1; Emscripten; Ghidra; PyTorch 2.2.1; Python; Tree-sitter; Ubuntu 22.04.3 LTS; Wasm2c\rule[-1.2ex]{0pt}{0pt}} \\ \hline

\rowcolor[HTML]{FFFFFF}
45 & Tan et al., 2024 \cite{tan2024llm4decompile} & Y & Binary decompilation & \wrapcell{0.115\textwidth}{LLM4Decompile-End-1.3/6.7/33B \\ *DeepSeek-Coder-6.7B \\ *GPT-4o\rule[-1.2ex]{0pt}{0pt}} & Assembly code or Ghidra pseudo-code & \wrapcell{0.115\textwidth}{*EXEBENCH \\ *HumanEval-Decompile} & \conditions{\textbf{HW:} Eight NVIDIA A100-80GB GPUs \\ \textbf{SW:} GCC; Ghidra; LLaMA-Factory; Linux; Objdump; PyTorch; Python; vLLM} \\ \hline

\rowcolor[HTML]{F6F8F9}
46 & \conditions{Walton et al., \\2024 \cite{walton2024exploring}} & N & Android malware semantic analysis & GPT-4o-mini & Decompiled Java source code from Android Package Kit & Benchmark from VirusTotal and Similarweb & \conditions{\textbf{HW:} - \\ \textbf{SW:} CFR Decompiler; Dex2Jar; Python} \\ \hline

\rowcolor[HTML]{FFFFFF}
47 & Xie et al., 2024 \cite{xie2024resym} & Y & Local variable and data structure name/type recovery & \wrapcell{0.115\textwidth}{FieldDecoder/VarDecoder (StarCoder 3B-based) \\ *GPT-4} & Decompiled code from compiled binaries of C/C++ code & Dataset from 7,416 Github C/C++ projects & \conditions{\textbf{HW:} Four A100 GPUs \\ \textbf{SW:} IDA Pro; Clang; GHCC; Prolog-based reasoning system} \\ \hline

\rowcolor[HTML]{F6F8F9}
48 & Zhou et al., 2024 \cite{zhou2024enhancing} & N & Database implementation vulnerability detection & \wrapcell{0.115\textwidth}{GPT \\ Machine language model} & Disassembled code, Decompiled pseudocode from C code & - & \conditions{\textbf{HW:} - \\\textbf{SW:} AES; database-crypt.dll; DirectoryMonitor; IDA Pro 7.0; SEAL library; SQL engine; Windows 10 virtual machine; X64DBG\rule[-1.2ex]{0pt}{0pt}} \\ \hline

\end{longtable}
\endgroup



\subsection{Intra-Dimensional Relationships}
\label{sec:intra_dimensional_relationships}
Although Sec~\ref{sec:taxonomy}: Taxonomy defines individual attributes within each axis, these attributes should not be interpreted as isolated or mutually exclusive categories. 
Many LLM-based RE studies combine multiple objectives, representations, methods, and validation strategies. 
Accordingly, the internal structure of each dimension reveals the trade-offs and complementarities that shape methodological choices in this emerging area.

\textbf{Motivation axis.}
The objectives within the motivation axis form a continuum from application-oriented engineering goals to knowledge-oriented exploration. 
Performance and robustness are engineering-driven objectives.
The former emphasizes improvements in effectiveness, efficiency, and task accuracy~\cite{ achamyeleh2026helios, YANG2026104468}, whereas the latter concerns stability, reliability, and resilience under challenging RE conditions. 
Discovery, by contrast, is exploratory and focuses on using LLMs to surface overlooked problems, identify underexplored methods, or formulate new hypotheses for RE investigation. 
Interpretability occupies an intermediate position, because it supports both engineering and exploratory aims. 
Making low-level program artifacts comprehensible to human analysts is important for building reliable RE tools, but it also enables new insights into program behavior. 
For example, a performance-oriented decompilation system may also improve the clarity of generated summaries, but a discovery-oriented study depends on interpretive techniques to make its findings intelligible~\cite{kirla2025large,wong2025decllm, arasteh2025trim}. 
Thus, this continuum does not imply a hierarchy among objectives.
Otherwise, it highlights complementary motivations that connect practical tool development with broader methodological inquiry.

\textbf{Representation axis.}
The targets within the representation axis describe an abstraction spectrum as well as different structural forms of target information. 
Raw bytes and assembly code preserve the closest connection to machine-level execution, but they impose substantial demands on models because semantic structure, control flow, data flow, and function boundaries must be inferred from low-level artifacts~\cite{she2024wadec,sha2025llasm}. 
At higher abstraction levels, decompiled code and source code provide more accessible, source-like representations that are better aligned with LLM reasoning, but they may introduce reconstruction artifacts or obscure the absence of original source code~\cite{tan2024llm4decompile}. 
Graph representations are orthogonal to this linear abstraction spectrum.
Instead of primarily increasing readability, they make control, call, and data dependencies explicit~\cite{incollection, su2025disco}. 
Natural language further shifts the representation toward analyst-facing descriptions, explanations, and task instructions, while Other captures task-specific labels or annotations introduced by diverse RE tasks~\cite{wei2025exposing, ExposingtheGuardrails}. 
These targets expose a central trade-off between fidelity to real-world RE conditions, compatibility with LLM reasoning, and the ability to support faithful human interpretation.

\textbf{Operation axis.}
The methods in the operation axis are frequently combined rather than used in isolation. 
Prompting commonly serves as the most lightweight mode of applying general-purpose LLMs and often functions as a baseline even in studies that also use other techniques. 
Fine-tuning increases task specialization by adapting models to RE-specific data, but it introduces stronger dependencies on dataset quality, training recipes, and computational resources~\cite{liu2025armqwen2, ouf2025reverse}. 
RAG provides external grounding through documentation, signatures, vulnerability knowledge, or other reference materials~\cite{YANG2026104468, boronat2025mdre, yao2025hdldebugger}, but agent-based analysis furtherly integrates LLMs into interactive workflows involving tools such as disassemblers, decompilers, debuggers, symbolic-execution engines, or sandboxes~\cite{xue2026record, huang2025malpre}. 
Data generation addresses data scarcity by using LLMs to synthesize, augment, or label RE data~\cite{arai2025staged, greiner2025generalization, 11391932}, and human involvement cuts across these methods through expert-crafted prompts, manual validation, feedback, or analyst-in-the-loop reasoning~\cite{fyrbiak2017hardware, ouf2025reverse, wei2025exposing}. 
These relationships form a methodological continuum from lightweight model use to workflow-level system integration. 
Recognizing this structure helps explain why reported performance may depend not only on the LLM itself, but also on the surrounding data, tools, prompts, and human expertise.

\textbf{Validation axis.}
The evaluational attributes are complementary forms of evidence, rather than substitutable evaluation choices. 
Automated metric scoring supports scalable and repeatable comparison, but it may fail to capture semantic correctness, functional equivalence, or analyst usability~\cite{sha2025llasm, app16062801}. 
Expert-based assessment provides contextual and security-relevant judgment, especially for interpretability-oriented tasks, but it is costly, potentially subjective, and difficult to reproduce without clear protocols~\cite{siala2025towards, hu2024degpt}. 
Ground-truth validation offers a more objective basis for comparison when authoritative labels, benchmarks, or executable reference checks are available, but such ground truth is often scarce or difficult to define for open-ended RE tasks~\cite{achamyeleh2026helios, zhang2025rebert}. 
A mature evaluation practice therefore requires triangulation across these attributes, for example by calibrating automated metrics against expert assessment, or anchoring expert judgments in reproducible ground-truth checks~\cite{hatahet2025generating, jiang2023nova,11391932}. 
Without such integration, progress in LLM-based RE may remain difficult to compare across studies and over time.

\textbf{Magnitude axis.}
The training scale within the magnitude axis form a scale continuum in model use, adaptation, and learning investment. 
Proof-of-concept or few-shot scale studies provide low-cost evidence of feasibility and are useful for early task exploration, but they usually offer limited evidence of generalizability or robustness~\cite{niang2025enhancing}. 
Fine-tuning scale represents an intermediate scale, where pretrained models are adapted to RE-specific data to improve task alignment, often at the cost of stronger dependence on dataset availability, annotation quality, and reporting completeness~\cite{dramko2025idioms}. 
Pretraining or massive scale rely on large-scale training or continued pretraining over code, binary, assembly, or security-related corpora, offering broader transfer potential, but raising concerns about computational cost, opaque training data, and reproducibility~\cite{jiang2023nova}. 
The relationship among these attributes indicates that scale alone does not establish progress. 
Larger training regimes may improve capability, but meaningful comparison still depends on transparent workflow, comparable tasks, and reliable evaluation.

\textbf{Reliability axis.}
The three data quality attributes in the reliability axis are interdependent and determine the empirical credibility of LLM-based RE studies. 
Source reliability provides the foundation by indicating whether datasets are traceable, accessible, and verifiable. 
Preprocessing assumption determines whether the transformation from raw artifacts to model-ready representations can be reconstructed and whether important semantic information may have been preserved, abstracted, or discarded. 
Ground-truth validation reliability assesses whether labels or reference answers are clearly defined, documented, and validated. 
A weakness in any one of these attributes can undermine the others.
A well-sourced dataset may still yield unreliable conclusions if preprocessing is opaque, and validated labels cannot fully compensate for missing provenance or undocumented transformations~\cite{duan2024sbcm, su2025disco, ouf2025reverse}. 
Thus, reliability is best understood as a chain of evidence linking data origin, data transformation, and label validity. 
Establishing consistency across all three attributes is essential for producing trustworthy, comparable, and reproducible evidence in LLM-based RE.

Overall, the intra-dimensional relationships show that the taxonomy is not merely a set of independent categories. 
Each axis encodes a set of intra trade-offs, and many attributes function as complements rather than alternatives. 
This internal structure provides the basis for the cross-dimensional analysis that follows, where objectives, representations, methods, evaluation strategies, training scale, and data quality are examined jointly.

\subsection{Cross-Dimensional Interactions and Insights}
\begin{figure*}[t]
    \centering
    \captionsetup[subfigure]{
        position=top,
        labelformat=parens,
        labelsep=space,
        justification=raggedright,
        singlelinecheck=false,
        font=footnotesize,
        labelfont=bf,
        skip=4pt
    }

    \begin{subfigure}[t]{0.64\textwidth}
        \caption{  Objective (V) - Method (H)}
        \label{fig:objective_methods}
        \includegraphics[
            width=\linewidth
        ]{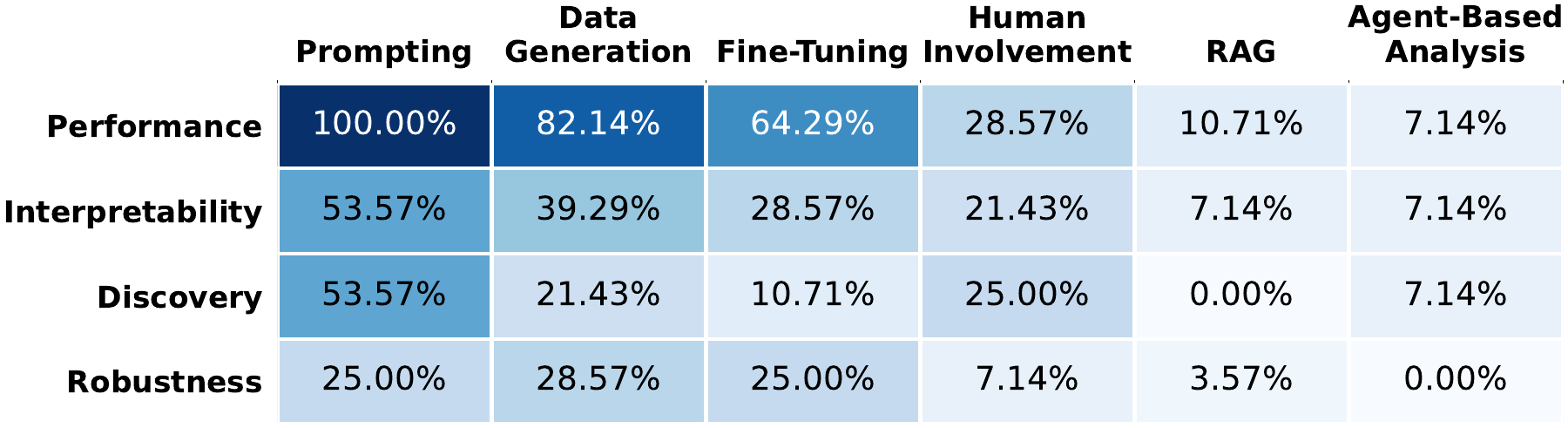}
    \end{subfigure}
    \hfill
    \begin{subfigure}[t]{0.355\textwidth}
        \caption{  Objective (V) - Evaluation (H)}
        \label{fig:objective_evaluation_metrics}
        \includegraphics[
            width=\linewidth
        ]{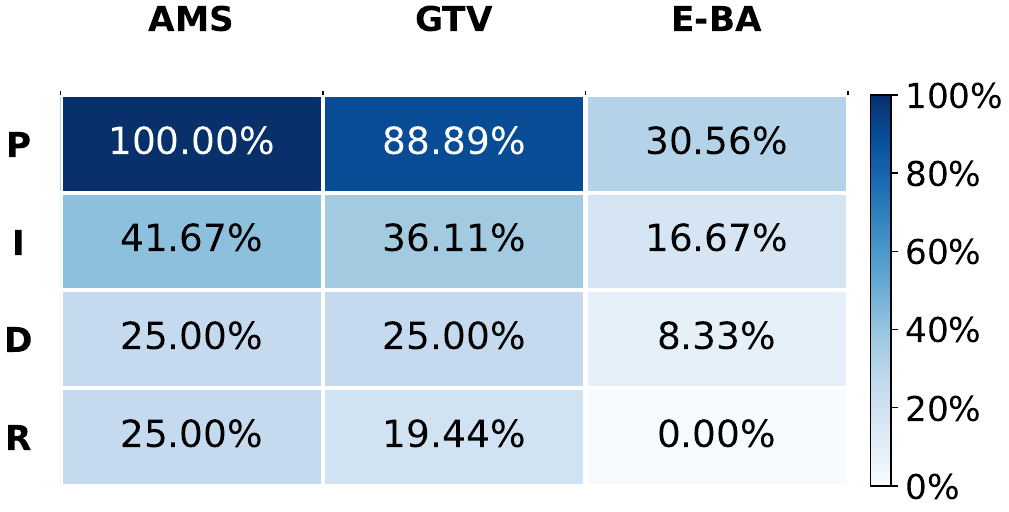}
    \end{subfigure}
    \hfill

    \caption{Min--max-normalized heatmaps illustrating the relationships between motivation-axis objectives and validation- or operation-axis attributes across the collected LLM-based RE studies. 
    (a) Motivation objectives are shown as rows and validation-axis evaluations are shown as columns. 
    (b) Motivation objectives are shown as rows and operation-axis methods are shown as columns. 
    V and H denote the vertical and horizontal attributes of each heatmap, respectively.}
    \label{fig:objective_heatmaps}
\end{figure*}

\begin{figure*}[t]
    \centering

    \captionsetup[subfigure]{
        position=top,
        labelformat=parens,
        labelsep=space,
        justification=raggedright,
        singlelinecheck=false,
        font=footnotesize,
        labelfont=bf,
        skip=4pt
    }

    \begin{subfigure}[t]{0.85\textwidth}
        \caption{Input Target (Vertical) - Method (Horizontal)}
        \label{fig:main_input_methods}
        \centering
        \includegraphics[width=\linewidth]
        {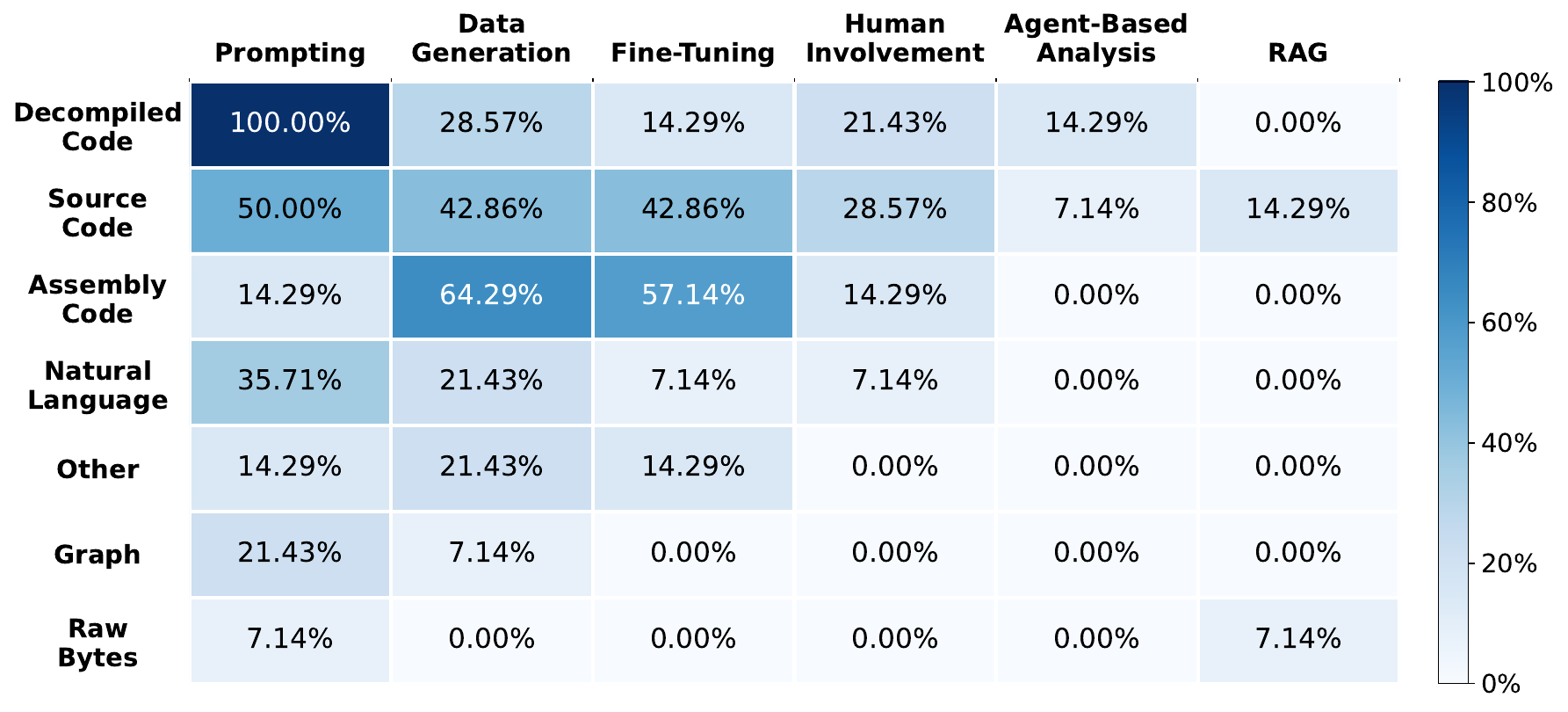}
    \end{subfigure}

    \vspace{0.8em}

    \begin{subfigure}[t]{0.85\textwidth}
        \caption{Output Target (Vertical) - Method (Horizontal)}
        \label{fig:main_output_methods}
        \centering
        \includegraphics[width=\linewidth]
        {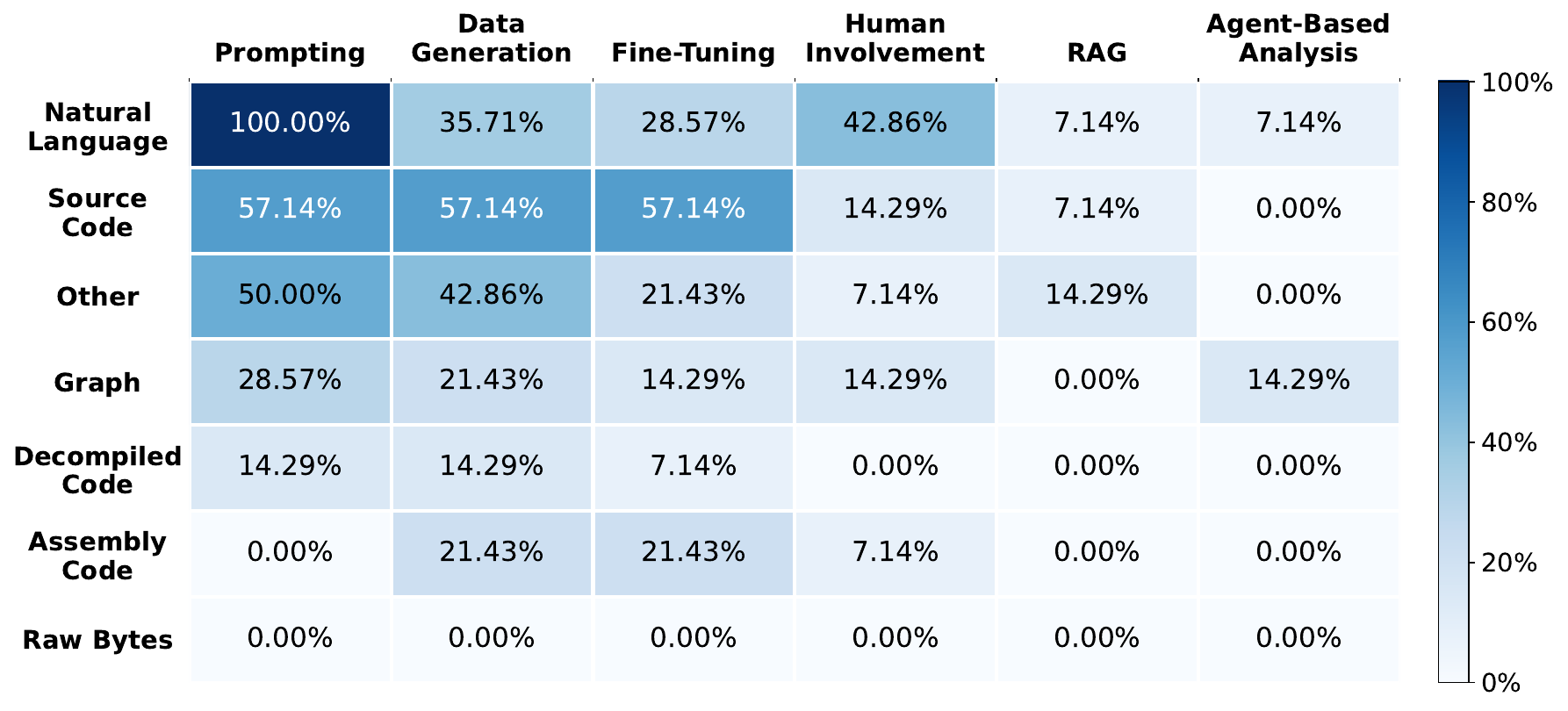}
    \end{subfigure}

    \caption{Min--max-normalized heatmaps illustrating the relationships between primary input/output targets in the representation axis and operation axis methods across the collected LLM-based RE studies: 
    (a) primary input targets as rows and methods as columns; 
    (b) primary output targets as rows and methods as columns.}
    \label{fig:output_target_method_heatmaps}
\vspace{-10pt}
\end{figure*}

Having examined the relationships among attributes within each taxonomy dimension, we now turn to the interactions across dimensions. 
The six dimensions of the proposed taxonomy are analytically distinct, but they are not independent in practice. 
Motivations shape the choice of representations and methods. 
Targets' representations constrain feasible methods and evaluation strategies. 
Validation results are meaningful only when interpreted together with methodological design, data scale, and reliability conditions. 
Figures~\ref{fig:objective_heatmaps}--\ref{fig:output_target_method_heatmaps} report min--max-normalized co-occurrence patterns among selected taxonomy dimensions. 
The cell values indicate relative association strength within each heatmap and should not be interpreted as raw frequencies, corpus-level percentages, or causal effects.

\textbf{Motivation, operation, and validation.}
Figure~\ref{fig:objective_heatmaps} shows the cross-dimensional relationships between motivation-axis objectives and operation- or validation-axis attributes. 
We first examine the motivation--operation interaction in Figure~\ref{fig:objective_heatmaps}(a). 
Prompting shows strong normalized associations across multiple objectives, reflecting its role as a lightweight and flexible mechanism for applying general-purpose LLMs to diverse RE tasks. 
Performance-oriented objectives are closely associated with prompting and data generation, suggesting that many studies use LLMs either as direct task solvers or as auxiliary tools for constructing RE data. 
Interpretability-oriented objectives show broader methodological associations, including prompting, fine-tuning, data generation, and human involvement, indicating the need for additional adaptation or analyst-facing support. 
By contrast, discovery-oriented studies remain largely prompt-driven, with limited association with RAG or agent-based analysis, while robustness-oriented studies show weaker methodological associations overall. 
These patterns suggest that methodological support is comparatively mature for performance-oriented work, but remains less developed for discovery and robustness.

We next examine the motivation--validation interaction in Figure~\ref{fig:objective_heatmaps}(b). 
Performance-oriented objectives show the strongest normalized associations with automated metric scoring and ground-truth validation, consistent with their emphasis on measurable improvements in accuracy, efficiency, or task success. 
However, their weaker association with expert-based assessment suggests that human-centered evaluation of semantic adequacy and analyst usability is less consistently incorporated. 
Interpretability, discovery, and robustness objectives show weaker associations with established validation strategies, indicating that explanations, exploratory findings, and reliability claims are still less systematically evaluated than performance claims. 
Overall, the heatmaps suggest that LLM-based RE has developed more mature methodological and validation practices for performance-oriented studies, while interpretability, discovery, and robustness still require stronger methodological grounding and evaluation protocols.

\begin{figure*}[t]
    \centering

    \captionsetup[subfigure]{
        position=top,
        labelformat=parens,
        labelsep=space,
        justification=raggedright,
        singlelinecheck=false,
        font=footnotesize,
        labelfont=bf,
        skip=4pt
    }

    \begin{subfigure}[t]{0.497\textwidth}
        \caption{Input Target (Vertical) - Evaluation (Horizontal)}
        \label{fig:main_input_metrics}
        \centering
        \includegraphics[
            width=\linewidth
        ]{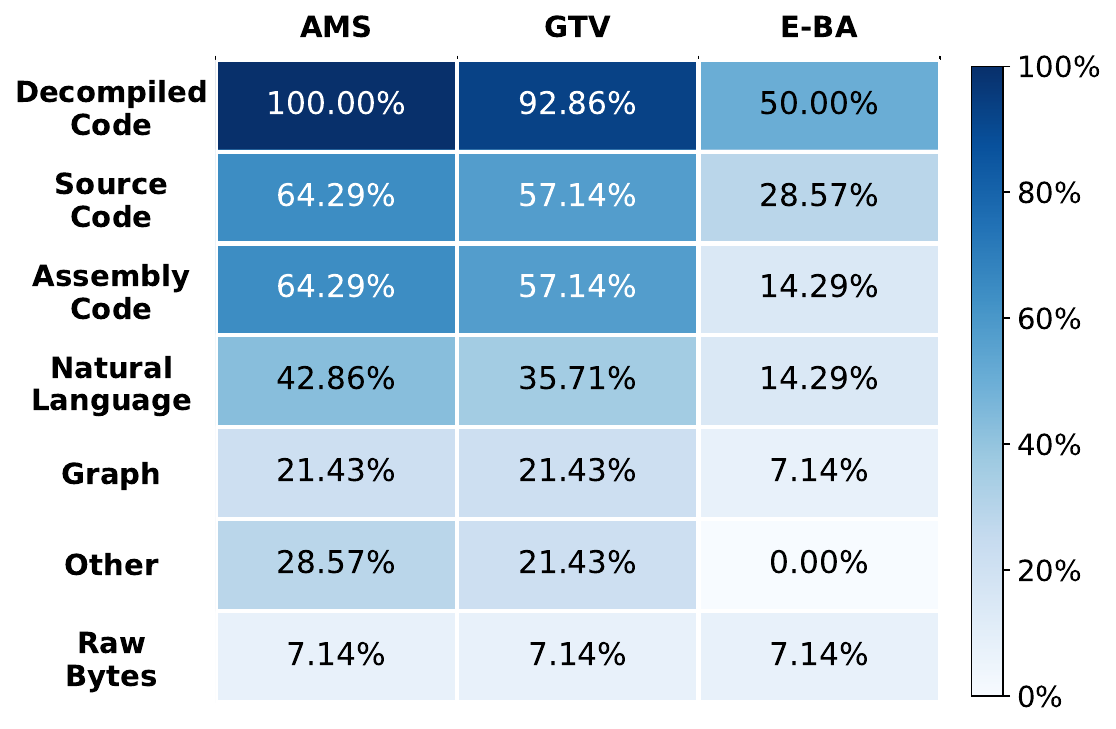}
    \end{subfigure}
    \hfill
    \begin{subfigure}[t]{0.497\textwidth}
        \caption{Output Target (Vertical) - Evaluation (Horizontal)}
        \label{fig:main_output_metrics}
        \centering
        \includegraphics[
            width=\linewidth
        ]{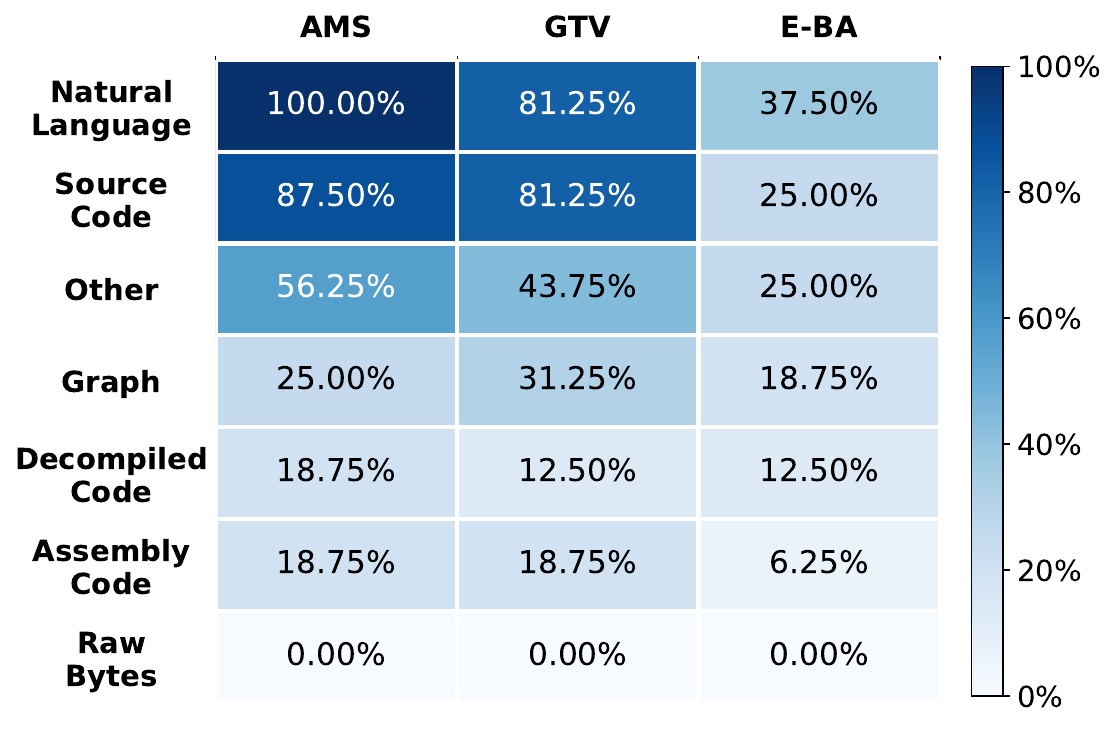}
    \end{subfigure}
    \caption{Min--max-normalized heatmaps illustrating the relationships between primary representation-axis input/output targets and validation-axis evaluations across the collected LLM-based RE studies. 
    (a) Primary input targets are shown as rows and evaluations are shown as columns. 
    (b) Primary output targets are shown as rows and evaluations are shown as columns. 
    }
    \label{fig:target_evaluation_heatmaps}
\end{figure*}

\textbf{Representation and operation.}
Figure~\ref{fig:output_target_method_heatmaps} presents min--max-normalized co-occurrence patterns between representation-axis targets and operation-axis methods. 
The values therefore indicate relative associations rather than corpus-level percentages. 
For input targets, decompiled code shows the strongest association with prompting, suggesting that source-like reconstructed code aligns well with the instruction-following and text-processing capabilities of current LLMs. 
Source code inputs are associated with prompting, data generation, and fine-tuning, indicating their use in both direct reasoning and model or data adaptation. 
In contrast, assembly code inputs are more strongly associated with data generation and fine-tuning, reflecting the additional adaptation needed for machine-level semantics. 
Raw bytes and graph-based inputs show weaker associations with most methods, suggesting that low-level and relational representations remain difficult to integrate into standard LLM workflows.
For output targets, natural language is most strongly associated with prompting and shows a notable association with human involvement, indicating the use of LLMs to produce explanations, summaries, and analyst-facing descriptions. 
Source code outputs are also associated with prompting, data generation, and fine-tuning, reflecting their role in code reconstruction and source-level generation. 
Overall, the representation--operation interaction reveals a methodological bias toward LLM-compatible textual representations, especially decompiled code as input and natural language or source code as output. 
This facilitates rapid progress, but also highlights a limitation: representations that are convenient for LLMs may not fully preserve machine-level semantics or relational program structure.

\textbf{Representation and validation.}
Figure~\ref{fig:target_evaluation_heatmaps} presents min--max-normalized co-occurrence patterns between representation-axis targets and validation-axis evaluations. 
The values indicate relative associations rather than corpus-level percentages. 
For input targets, decompiled code shows the strongest association with automated metric scoring and ground-truth validation, suggesting that source-like reconstructed code is easier to evaluate through metrics, labels, or reference outputs. 
Source and assembly code also show visible associations with these validation strategies, whereas raw bytes, graphs, and other specialized inputs remain more weakly connected to existing evaluation practices.
For output targets, natural language and source code show the strongest associations with automated metric scoring and ground-truth validation. 
However, natural-language outputs also raise a stronger need for expert-based assessment, since fluent explanations do not necessarily imply faithful recovery of program behavior. 
Overall, the representation--validation interaction suggests that evaluation is more mature for source-like and natural-language representations, but remains less standardized for low-level, graph-based, or structurally constrained artifacts. 
This highlights the need for representation-aware metrics, verifiable ground truth, and expert judgment when semantic faithfulness and analyst usability are central.

\textbf{Implications across the six-dimensional taxonomy.}
Although the heatmaps primarily visualize interactions among motivation, representation, operation, and validation, their implications must be interpreted together with the magnitude and reliability axes. 
The observed cross-dimensional patterns suggest that current LLM-based RE research is shaped by a strong alignment among three factors: LLM-compatible representations, lightweight operational methods, and scalable validation strategies. 
In particular, decompiled code and natural language are frequently aligned with prompting, data generation, automated metric scoring, and ground-truth validation. 
This alignment has enabled rapid exploration of LLM capabilities in RE, because these representations and methods fit the text-centric interface of existing LLMs and can be evaluated with comparatively accessible metrics or reference labels.
However, this alignment also exposes a structural limitation. 
The settings that are easiest for LLMs to process and evaluate are not necessarily those that best capture the distinctive challenges of RE. 
Low-level artifacts, graph-structured program information, dynamic behaviors, obfuscation, platform-specific execution contexts, and analyst-in-the-loop reasoning remain harder to integrate into standard LLM workflows and harder to validate using conventional metrics. 
As a result, progress measured under LLM-friendly representations may not fully reflect progress on machine-level semantic grounding, realistic RE workflows, or security-critical deployment conditions.

The magnitude and reliability axes further qualify these findings. 
Prompting over decompiled or source-like representations can provide useful feasibility evidence, especially at proof-of-concept or few-shot scale, but stronger empirical claims require reliable datasets, transparent preprocessing, validated labels, and reproducible implementation settings. 
Similarly, fine-tuning, RAG, agent-based analysis, and data generation may improve specialization or workflow integration, but they also increase dependence on training data quality, toolchain configuration, benchmark design, and reporting completeness. 
Without sufficient reliability controls, improvements attributed to LLM capability may instead reflect dataset bias, preprocessing choices, expert intervention, or evaluation artifacts.

Overall, the cross-dimensional analysis indicates that LLM-based RE should not be evaluated along any single dimension in isolation. 
A study's contribution depends on the alignment among its motivation, representation, method, validation strategy, training scale, and data reliability evidence. 
So future work should move toward integrated evaluation practices that connect low-level program semantics with human-interpretable outputs, combine automated metrics with expert judgment where appropriate, and report data construction, preprocessing, model settings, and experimental environments in detail. 
Such coordinated progress is necessary for transforming LLM-based RE from a collection of promising task-specific demonstrations into a cumulative, comparable, and scientifically grounded research area.

\subsection{Qualitative Insights}
\label{quanlitative_insights}
Building on the quantitative landscape and intra/cross-dimensional analysis, we synthesize the qualitative patterns that shape the development of LLM-based RE. 
Rather than restating attribute distributions, the goal is to interpret why certain research configurations dominate, what trade-offs they introduce, and how they affect the maturity, reliability, and practical adoption of the field.

\textbf{From feasibility demonstrations to workflow-aware systems.}
The technical trajectory of LLM-based RE suggests a gradual shift from lightweight feasibility studies toward more integrated analysis workflows. 
Early studies primarily relied on direct prompting and data generation of code, assembly, or binary artifacts~\cite{pearce2022pop}. 
More recent work increasingly explores fine-tuning, RAG, human involvement  and agent-based analysis, often integrating external tools such as decompilers, debuggers, symbolic execution engines, or domain knowledge bases~\cite{yao2025hdldebugger,zhou2024enhancing}. 
This evolution reflects a broader movement from asking whether general-purpose LLMs can support RE tasks toward designing systems that can operate within realistic RE workflows. 
However, the shift also increases the dependence on toolchains, data construction, model configurations, and interaction protocols, making reproducibility more difficult unless these components are explicitly reported.

\textbf{LLM compatibility versus machine-level fidelity.}
A central qualitative tension concerns the mismatch between representations that are convenient for LLMs and representations that preserve the distinctive challenges of RE. 
Text-like artifacts, especially decompiled code, source-like representations, and natural-language outputs, align well with current LLM interfaces and therefore, enable rapid progress through prompting and generation. 
By contrast, raw bytes, graph structures, dynamic traces, and obfuscated artifacts are closer to real-world RE conditions, but harder to encode, reason over, and evaluate. 
This creates a methodological bias: studies may obtain promising results on LLM-friendly representations, while leaving harder questions of machine-level semantic grounding underexplored. 
Future systems therefore need stronger bridges between low-level execution semantics, relational program structure, and human-interpretable explanations.

\textbf{Evaluation proxies and reproducibility bottlenecks.}
Current evaluation practices often rely on scalable metrics, reference labels, or task-specific benchmarks, but these proxies do not always capture semantic correctness, analyst usefulness, or security relevance. 
For example, a generated decompilation output may appear syntactically plausible or textually similar to a reference, while still misrepresenting control flow, data dependencies, or program behavior~\cite{achamyeleh2026helios}. 
Similarly, objectives optimized for compilability, label accuracy, or benchmark scores may encourage models to improve proxy performance, without necessarily improving faithful RE reasoning. 
This issue is amplified by uneven reporting of benchmarks, model comparisons, open-source artifacts, preprocessing pipelines, hardware/software environments, and ground-truth validation. 
As shown in the quantitative analysis, standard benchmarks, comparative model evaluation, and open-source resources are not yet consistently adopted across the field. 
Consequently, the field needs evaluation protocols that combine representation-aware metrics, executable or semantic checks, expert assessment, and transparent reporting of experimental conditions.

\textbf{Human analysts as hidden variables and collaborators.}
Human expertise remains central to LLM-based RE, but this role is difficult to isolate. 
Analysts may design prompts, select examples, validate outputs, interpret ambiguous results, or decide when to trust or override model suggestions. 
When such involvement is underreported, it becomes difficult to distinguish model capability from analyst expertise. 
At the same time, human involvement should not be treated merely as a source of evaluation cost. 
In realistic RE settings, analysts and LLMs may form interactive workflows in which the model proposes hypotheses, explains artifacts, retrieves evidence, or operates tools, while the analyst provides domain judgment and security context. 
The design of this interaction is critical.
Explanations, uncertainty signals, retrieved evidence, and transparent reasoning traces may shape user trust, error detection, and decision revision~\cite{white2023mixed,leong2024exploratory}. 
These human-centered dimensions remain underexplored, but are essential for high-stakes RE applications.

\textbf{Domain-driven forces and future maturation.}
The development of LLM-based RE is shaped not only by advances in model architecture, but also by the domain-specific pressures of cybersecurity. 
The scarcity of labeled RE data motivates data generation and weak supervision~\cite{arai2025staged, choi2024chatdeob}. 
The complexity of binaries and execution environments motivates tool-augmented and agent-based workflows~\cite{brown2025large}.
The security-critical nature of RE motivates stronger attention to reliability, interpretability, and dual-use risks~\cite{11391932, hatahet2025generating}. 
Similar maturation patterns have appeared in adjacent security areas, including vulnerability analysis, fuzzing, program repair, and intrusion detection, where early proof-of-concept demonstrations were followed by task-specific adaptation, benchmark construction, and hybrid symbolic--neural systems~\cite{zhou2024enhancing, oliinyk2024fuzzing, app16062801}. 
For LLM-based RE, this suggests that the next stage of progress will depend less on isolated model improvements and more on integrated research infrastructure: reliable benchmarks, transparent datasets, reproducible toolchains, human-centered evaluation, and mechanisms for grounding high-level explanations in machine-level program behavior.

Overall, the qualitative insights indicate that LLM-based RE is transitioning from opportunistic use of general-purpose models toward a more structured research area. 
The main challenge is no longer only whether LLMs can produce useful outputs for individual RE tasks, but whether such outputs can be trusted, reproduced, compared, and integrated into realistic security workflows. 
Addressing this challenge requires coordinated progress across model design, representation learning, evaluation methodology, data reliability, and analyst-centered system design.

\section{Gaps and Future Research Agenda}
\label{sec:gaps_future_agenda}
The preceding analyses show that LLM-based RE is an active but still consolidating research area. 
Existing studies have demonstrated the potential of LLMs for decompilation, binary understanding, vulnerability analysis, malware analysis, semantic recovery, and related RE tasks. 
However, the evidence also reveals several structural gaps that limit comparability, reproducibility, semantic grounding, and practical adoption. 
This section summarizes these gaps and outlines future directions for developing more reliable, interpretable, and security-relevant LLM-based RE systems.

\textbf{Gap 1: Performance-centric objectives and limited exploration of broader research goals.}
Current LLM-based RE research remains strongly performance-oriented. 
This focus has accelerated progress on task accuracy, efficiency, and automation, but it also risks narrowing the field toward benchmark-driven optimization. 
Interpretability, robustness, and discovery remain less systematically studied, even though they are central to security-critical RE. 
Interpretability is needed to determine whether LLM outputs are understandable and faithful to program behavior; robustness is needed to assess stability under obfuscation, noise, prompt perturbation, and platform variation; and discovery is needed to support exploratory analysis of overlooked RE problems and undocumented behaviors.

\textit{Future agenda.}
Future work should broaden the objective space beyond performance optimization. 
Interpretability-oriented studies should evaluate whether LLMs can generate faithful high-level explanations grounded in low-level program semantics. 
Robustness-oriented work should introduce stress tests involving obfuscation, packing, compiler diversity, architecture variation, adversarial prompts, and toolchain instability. 
Discovery-oriented studies should examine how LLMs can help identify neglected RE tasks, analysis targets, and hypotheses for further investigation.

\textbf{Gap 2: Overreliance on LLM-friendly representations.}
Existing studies often favor decompiled code, source-like artifacts, and natural-language outputs because these representations align well with current LLM interfaces. 
While this choice enables rapid progress, it does not fully capture the distinctive difficulty of RE, where analysts often work with raw bytes, assembly code, graph structures, dynamic traces, obfuscated binaries, and platform-specific execution contexts~\cite{YANG2026104468, greiner2025generalization, liu2025armqwen2}. 
As a result, progress on text-like representations may not directly translate into progress on machine-level semantic grounding.

\textit{Future agenda.}
Future research should develop representation strategies that connect low-level artifacts with high-level reasoning. 
Promising directions include hybrid representations that combine assembly, decompiled code, control-flow graphs, data-flow graphs, execution traces, and natural-language context. 
LLM-based RE systems should also be evaluated for cross-representation consistency, such as whether a generated explanation, recovered source code, and graph abstraction remain faithful to the original binary behavior.

\textbf{Gap 3: Methodological maturity remains uneven.}
Prompting and data generation are widely used because they are accessible and require limited infrastructure. 
However, these methods often provide limited guarantees of robustness, or reproducibility~\cite{liao2025augmenting, ma2024one}. 
More specialized approaches, including fine-tuning, retrieval-augmented generation, and agent-based analysis, can improve domain alignment and workflow integration, but they introduce new dependencies on training data, retrieval corpora, toolchains, orchestration logic, and implementation environments.

\textit{Future agenda.}
Future studies should move toward methodologically grounded hybrid pipelines. 
Fine-tuning should be paired with transparent dataset construction and reproducible training recipes. 
RAG should rely on reliable, versioned, and task-relevant knowledge sources. 
Agent-based analysis should report toolchain configurations, intermediate states, decision procedures, and failure cases. 
Data generation should include validation mechanisms to prevent synthetic errors or label noise from propagating into downstream training and evaluation.

\textbf{Gap 4: Evaluation, data quality, and reproducibility are not yet standardized.}
Current evaluations frequently rely on automated metrics and ground-truth validation, while expert-based assessment is less consistently incorporated. 
Automated metrics support scalability, but may not capture functional equivalence, semantic faithfulness, analyst usability, or security relevance. 
Meanwhile, reproducibility is limited by uneven reporting of benchmarks, preprocessing pipelines, label definitions, model versions, prompts, hardware/software settings, and open-source artifacts~\cite{zhou2024enhancing, li-klabjan-2025-reverse, yao2025hdldebugger}.

\textit{Future agenda.}
The field needs standardized and representation-aware evaluation protocols. 
For source-like outputs, evaluations should combine syntactic metrics with executable, semantic, or behavioral checks. 
For natural-language explanations, they should assess faithfulness, completeness, specificity, and analyst usefulness. 
For low-level or graph-based artifacts, validation should account for architecture-specific semantics, control-flow recovery, structural correctness, and execution behavior. 
Future studies should also report dataset provenance, preprocessing steps, tool versions, parameter settings, label validation procedures, model configurations, and experimental environments. 
When datasets cannot be released, authors should provide metadata, construction procedures, representative examples, and reproducible scripts where feasible.

\textbf{Gap 5: Human involvement and security risks remain under-specified.}
Human analysts play a central role in LLM-based RE by designing prompts, selecting examples, validating outputs, interpreting ambiguous results, and deciding whether to trust or override model suggestions. 
When such involvement is underreported, it becomes difficult to separate model capability from analyst expertise. 
At the same time, LLM-based RE is inherently dual-use: the same capabilities that support defensive analysis may also assist exploit development, malware adaptation, or evasion of analysis tools~\cite{walton2024exploring, zhou2024enhancing}. 
Additional risks arise from hallucination, prompt injection, unsafe tool invocation, sensitive artifact leakage, and dependence on closed or changing proprietary models~\cite{sriramanan2024llm, greshake2023not,awal2025impedance}.

\textit{Future agenda.}
Future work could model and report human involvement, including expert-crafted prompts, manual annotation, output validation, feedback-based refinement, and analyst-in-the-loop reasoning. 
Evaluation should report annotator expertise, judgment criteria, disagreement resolution, and analyst effort. 
Responsible deployment requires safeguards for untrusted inputs, tool-mediated execution, sensitive artifacts, and model-output verification. 
Risk analysis should be integrated into benchmark design and system evaluation, rather than treated only as a discussion point.

\textbf{Summary.}
These gaps suggest that the next phase of LLM-based RE should move from isolated task demonstrations toward integrated, reproducible, and security-grounded research practices. 
Progress will depend on coordinated development across objectives, representations, methods, evaluation protocols, training scale, data reliability, and human-centered system design. 
Addressing these challenges can help transform LLM-based RE from a rapidly expanding collection of promising studies into a cumulative, comparable, and scientifically grounded research area.

\section{Conclusion}
This survey provides a systematic review of 48 peer-reviewed studies on LLM-based RE. 
To organize this emerging and fragmented literature, we develop a six-dimensional taxonomy covering motivation, representation, operation, validation, magnitude, and reliability. 
Using this taxonomy, we characterize how LLMs have been applied to RE tasks, how program artifacts are represented, how methods are designed and evaluated, and how data quality and reproducibility affect empirical claims.
Our analysis shows that LLM-based RE has progressed rapidly, but remains unbalanced in methodological maturity and empirical rigor. 
Current studies are largely performance-driven, they often rely on LLM-compatible representations, such as decompiled code and natural-language outputs, and frequently use prompting or data generation as accessible operational strategies. 
At the same time, challenges remain in semantic grounding, benchmark standardization, expert-centered evaluation, data and preprocessing transparency, human-involvement reporting, and reproducible implementation settings.

Future progress requires moving beyond isolated task demonstrations toward integrated and verifiable research practices. 
Important directions include connecting low-level program semantics with faithful human-interpretable explanations, developing representation-aware evaluation protocols, establishing reliable and shareable benchmarks, improving transparency in datasets and toolchains, and addressing dual-use and deployment risks. 
By consolidating existing work and clarifying cross-dimensional dependencies, this survey provides a foundation for more cumulative, comparable, and trustworthy research on LLM-based RE.

\bibliographystyle{ACM-Reference-Format}
\bibliography{main}


\clearpage
\appendix
\section{Appendix}

\begin{center}
\centering
\captionof{table}{Percentage distribution of 48 published LLM-based RE studies by objective from 2024 to 2026.}
\label{tab:objective-distribution}
\small
\setlength{\tabcolsep}{0pt}
\renewcommand{\arraystretch}{1.15}

\begin{tabular}{%
>{\centering\arraybackslash}p{0.20\textwidth}|
>{\centering\arraybackslash}p{0.20\textwidth}|
>{\centering\arraybackslash}p{0.20\textwidth}|
>{\centering\arraybackslash}p{0.20\textwidth}|
>{\centering\arraybackslash}p{0.20\textwidth}}
\toprule
\toprule
\textbf{Year $\backslash$ Obj} &
\textbf{Discovery} &
\textbf{Interpretability} &
\textbf{Performance} &
\textbf{Robustness} \\
\midrule
\textbf{2026}  & 6.25\%  & 8.33\%  & 12.50\% & 4.17\%  \\
\textbf{2025}  & 20.83\% & 25.00\% & 60.42\% & 20.83\% \\
\textbf{2024}  & 6.25\%  & 10.42\% & 16.67\% & 4.17\%  \\
\textbf{Total} & 33.33\% & 43.75\% & 89.58\% & 29.17\% \\
\bottomrule
\bottomrule
\end{tabular}
\vspace{-5pt}
\end{center}
\vspace{10pt}

\begin{center}
\centering
\captionof{table}{Percentage distribution of 48 published LLM-based RE studies by target from 2024 to 2026.}
\label{tab:target-distribution}
\small
\setlength{\tabcolsep}{6pt}
\renewcommand{\arraystretch}{1.15}

\resizebox{\textwidth}{!}{%
\begin{tabular}{c|c|c|c|c|c|c|c|c}
\toprule
\toprule
\textbf{Target Position} &
\textbf{Year $\backslash$ Tar} &
\textbf{Assembly Code} &
\textbf{Decompiled Code} &
\textbf{Graph} &
\textbf{Natural Language} &
\textbf{Raw Bytes} &
\textbf{Source Code} &
\textbf{Other} \\
\midrule

\multirow{4}{*}{\textbf{Input}}
& 2026  & -       & 6.25\%  & 2.08\% & -       & 2.08\% & 4.17\%  & -       \\
& 2025  & 14.58\% & 20.83\% & 2.08\% & 8.33\%  & -      & 16.67\% & 6.25\%  \\
& 2024  & 6.25\%  & 6.25\%  & 2.08\% & 4.17\%  & -      & -       & 2.08\%  \\
& \textbf{Total} & 20.83\% & 33.33\% & 6.25\% & 12.50\% & 2.08\% & 20.83\% & 8.33\%  \\
\midrule

\multirow{4}{*}{\textbf{Output}}
& 2026  & -       & -       & 2.08\% & 8.33\%  & -      & 2.08\%  & 4.17\%  \\
& 2025  & 6.25\%  & 4.17\%  & 6.25\% & 20.83\% & -      & 20.83\% & 14.58\% \\
& 2024  & -       & 2.08\%  & 2.08\% & 6.25\%  & -      & 8.33\%  & 2.08\%  \\
& \textbf{Total} & 6.25\% & 6.25\% & 10.42\% & 35.42\% & - & 31.25\% & 20.83\% \\

\bottomrule
\bottomrule
\end{tabular}}
\vspace{-5pt}
\end{center}
\vspace{10pt}
\begin{center}
\centering
\captionof{table}{Percentage distribution of 48 published LLM-based RE studies by method from 2024 to 2026.}
\label{tab:method-distribution}
\small
\setlength{\tabcolsep}{6pt}
\renewcommand{\arraystretch}{1.15}

\resizebox{\textwidth}{!}{%
\begin{tabular}{c|c|c|c|c|c|c}
\toprule
\toprule
\textbf{Year $\backslash$ Met} &
\textbf{Agent-Based Analysis} &
\textbf{Data Generation} &
\textbf{Fine-Tuning} &
\textbf{Human Involvement} &
\textbf{Prompting} &
\textbf{RAG} \\
\midrule
\textbf{2026}  & 2.08\% & 4.17\%  & 2.08\%  & 6.25\%  & 12.50\% & 2.08\% \\
\textbf{2025}  & 4.17\% & 33.33\% & 31.25\% & 12.50\% & 41.67\% & 4.17\% \\
\textbf{2024}  & -      & 12.50\% & 6.25\%  & 2.08\%  & 12.50\% & -      \\
\textbf{Total} & 6.25\% & 50.00\% & 39.58\% & 20.83\% & 66.67\% & 6.25\% \\
\bottomrule
\bottomrule
\end{tabular}%
}
\end{center}
\vspace{10pt}
\begin{center}
\centering
\small
\setlength{\tabcolsep}{0pt}
\renewcommand{\arraystretch}{1.15}

\begin{minipage}[t]{0.49\textwidth}
\centering
\captionof{table}{Percentage distribution of 48 published LLM-based RE studies by evaluation from 2024 to 2026.}
\label{tab:evaluation-distribution}

\begin{tabular}{%
>{\centering\arraybackslash}p{0.25\linewidth}|
>{\centering\arraybackslash}p{0.25\linewidth}|
>{\centering\arraybackslash}p{0.25\linewidth}|
>{\centering\arraybackslash}p{0.25\linewidth}}
\toprule
\toprule
\textbf{Year $\backslash$ Eva} &
\textbf{AMS} &
\textbf{E-BA} &
\textbf{GTV} \\
\midrule
\textbf{2026}  & 12.50\% & 6.25\%  & 8.33\%  \\
\textbf{2025}  & 62.50\% & 25.00\% & 58.33\% \\
\textbf{2024}  & 16.67\% & 4.17\%  & 14.58\% \\
\textbf{Total} & 91.67\% & 35.42\% & 81.25\% \\
\bottomrule
\bottomrule
\end{tabular}
\end{minipage}
\hfill
\begin{minipage}[t]{0.49\textwidth}
\centering
\captionof{table}{Percentage distribution of 48 published LLM-based RE studies by training scale from 2024 to 2026.}
\label{tab:training-scale-distribution}

\begin{tabular}{%
>{\centering\arraybackslash}p{0.25\linewidth}|
>{\centering\arraybackslash}p{0.25\linewidth}|
>{\centering\arraybackslash}p{0.25\linewidth}|
>{\centering\arraybackslash}p{0.25\linewidth}}
\toprule
\toprule
\textbf{Year $\backslash$ TS} &
\textbf{FTS} &
\textbf{PC/FS} &
\textbf{PT/MS} \\
\midrule
\textbf{2026}  & 2.08\%  & 10.42\% & -      \\
\textbf{2025}  & 29.17\% & 37.50\% & 2.08\% \\
\textbf{2024}  & 6.25\%  & 12.50\% & -      \\
\textbf{Total} & 37.50\% & 60.42\% & 2.08\% \\
\bottomrule
\bottomrule
\end{tabular}
\end{minipage}
\end{center}

\vspace{10pt}

\clearpage
\begin{center}
\captionof{table}{Percentage distribution of 48 published LLM-based RE studies by data quality from 2024 to 2026.}
\label{tab:data-quality-distribution}
\small
\setlength{\tabcolsep}{2pt}
\renewcommand{\arraystretch}{1.15}

\begin{tabular}{%
>{\centering\arraybackslash}p{0.10\textwidth}|
>{\centering\arraybackslash}p{0.09\textwidth}
>{\centering\arraybackslash}p{0.09\textwidth}
>{\centering\arraybackslash}p{0.09\textwidth}|
>{\centering\arraybackslash}p{0.09\textwidth}
>{\centering\arraybackslash}p{0.09\textwidth}
>{\centering\arraybackslash}p{0.09\textwidth}|
>{\centering\arraybackslash}p{0.09\textwidth}
>{\centering\arraybackslash}p{0.09\textwidth}
>{\centering\arraybackslash}p{0.07\textwidth}}
\toprule
\toprule
\textbf{Year $\backslash$ DQ} &
\multicolumn{3}{c|}{\textbf{GTV Reliability}} &
\multicolumn{3}{c|}{\textbf{Preprocessing Assumption}} &
\multicolumn{3}{c}{\textbf{Source Reliability}} \\
\cmidrule(lr){2-4}
\cmidrule(lr){5-7}
\cmidrule(lr){8-10}
&
\textbf{H} & \textbf{M} & \textbf{-} &
\textbf{H} & \textbf{M} & \textbf{-} &
\textbf{H} & \textbf{M} & \textbf{-} \\
\midrule
\textbf{2026}  & 8.33\%  & 4.17\%  & -      & 2.08\%  & 8.33\%  & 2.08\%  & 6.25\%  & 6.25\%  & -      \\
\textbf{2025}  & 54.17\% & 12.50\% & 2.08\% & 6.25\%  & 45.83\% & 16.67\% & 35.42\% & 33.33\% & -      \\
\textbf{2024}  & 12.50\% & 2.08\%  & 4.17\% & 8.33\%  & 6.25\%  & 4.17\%  & 6.25\%  & 10.42\% & 2.08\% \\
\textbf{Total} & 75.00\% & 18.75\% & 6.25\% & 16.67\% & 60.42\% & 22.92\% & 47.92\% & 50.00\% & 2.08\% \\
\bottomrule
\bottomrule
\end{tabular}
\vspace{-5pt}
\end{center}


\end{document}